\documentclass[fleqn,usenatbib]{mnras}

\usepackage{newtxtext,newtxmath}

\usepackage[T1]{fontenc}

\DeclareRobustCommand{\VAN}[3]{#2}
\let\VANthebibliography\thebibliography
\def\thebibliography{\DeclareRobustCommand{\VAN}[3]{##3}\VANthebibliography}

\usepackage{graphicx}	
\usepackage{amsmath}	

\newcommand\brg{Br$\gamma$ }
\newcommand\ks{$K_s$ }

\title[Extreme intermediate-mass binaries]{A low-mass companion desert among intermediate-mass visual binaries: \\
The scaled-up counterpart to the brown dwarf desert}

\author[G. Duch\^ene et al.]{
Gaspard Duch\^ene,$^{1,2}$\thanks{E-mail: gduchene@berkeley.edu}
Jner Tzern Oon,$^{3,1}$
Robert J. De Rosa,$^{4}$
Patrick Kantorski,$^{1}$
Brandon Coy,$^{1,5}$
\newauthor 
Jason J. Wang,$^{6,7}$
Sandrine Thomas,$^{8}$
Jenny Patience,$^{9}$
Laurent Pueyo,$^{10}$
Eric L. Nielsen,$^{11}$
and Quinn Konopacky$^{12}$
\\
$^{1}$Astronomy Department, University of California Berkeley, Berkeley CA 94720-3411, USA\\
$^{2}$Univ. Grenoble Alpes, CNRS, Institut d'Astrophysique et de Plan\'etologie de Grenoble, 38000 Grenoble, France\\
$^{3}$Department of Physics, University of Maryland, College Park, Maryland 20742, USA\\
$^{4}$European Southern Observatory, Alonso de C\'{o}rdova 3107, Vitacura, Santiago, Chile\\
$^{5}$Department of Earth, Planetary, and Space Sciences, University of California at Los Angeles, Los Angeles, 90095, USA\\
$^{6}$Department of Astronomy, California Institute of Technology, Pasadena, CA 91125, USA\\
$^{7}$Center for Interdisciplinary Exploration and Research in Astrophysics (CIERA) and Department of Physics and Astronomy, Northwestern University, Evanston, IL 60208, USA\\
$^{8}$Vera C. Rubin Observatory, 950 N Cherry Ave, Tucson AZ, 85719, USA\\
$^{9}$School of Earth and Space Exploration, Arizona State University, PO Box 871404, Tempe, AZ 85287, USA\\
$^{10}$Space Telescope Science Institute, 3700 San Martin Drive, Baltimore, MD 21218, USA\\
$^{11}$Department of Astronomy, New Mexico State University, 1320 Frenger Mall, Las Cruces, NM 88003-8001, USA\\
$^{12}$Center for Astrophysics and Space Sciences, University of California, San Diego, La Jolla, CA 92093, USA
}

\date{Accepted 2022 November 21. Received 2022 November 4; in original form 2022 July 29}

\pubyear{2022}

\begin{document}
\label{firstpage}
\pagerange{\pageref{firstpage}--\pageref{lastpage}}
\maketitle

\begin{abstract}
We present a high-contrast imaging survey of intermediate-mass (1.75--4.5\,$M_\odot$) stars to search for the most extreme stellar binaries, i.e., for the lowest mass stellar companions. Using adaptive optics at the Lick and Gemini observatories, we observed 169 stars and detected 24 candidates companions, 16 of which are newly discovered and all but three are likely or confirmed physical companions. Despite obtaining sensitivity down to the substellar limit for 75\% of our sample, we do not detect any companion below 0.3\,$M_\odot$, strongly suggesting that the distribution of stellar companions is truncated at a mass ratio of $q_\mathrm{min} \gtrsim0.075$. Combining our results with known brown dwarf companions, we identify a low-mass companion desert to intermediate mass stars in the range $0.02\lesssim q \lesssim0.05$, which quantitatively matches the known brown dwarf desert among solar-type stars. We conclude that the formation mechanism for multiple systems operates in a largely scale-invariant manner and precludes the formation of extremely uneven systems, likely because the components of a proto-binary accrete most of their mass after the initial cloud fragmentation. Similarly, the mechanism to form "planetary" ($q \lesssim 0.02$) companions likely scales linearly with stellar mass, probably as a result of the correlation between the masses of stars and their protoplanetary disks. Finally, we predict the existence of a sizable population of brown dwarf companions to low-mass stars and of a rising population of planetary-mass objects towards $\approx 1\,M_\mathrm{Jup}$ around solar-type stars. Improvements on current instrumentation will test these predictions.
\end{abstract}




\section{Introduction} \label{sec:intro}

Multiplicity is an integral and ubiquitous aspect of the star formation process \citep[for a review, see][]{duc13}. The broad diversity of multiple systems suggests that there most likely are several processes responsible for the formation of different types of binaries. Given the low likelihood of dynamical capture and the prevalence of dense cores hosting multiple protostars \citep{Sadavoy2017}, it is generally assumed that visual binaries, with separations $\gtrsim10$\,au, result from the fragmentation of a prestellar core prior to, or during, its initial collapse \citep{toh02}. The properties of multiple systems and their dependency on stellar mass, age and environment encode the physics at play during star formation. From the theoretical perspective, understanding how the initial fragmentation and subsequent accretion on the stellar seeds relate to the final properties of the multiple system is an ongoing effort \citep[for a recent review, see][]{Offner2022}.

Unlike the binary fraction or the distributions of semi-major axis and eccentricity, the distribution of mass-ratios ($q$, with $q\leq1$) is robust against subsequent dynamical evolution \citep{par13}, making it a pristine relic of the star formation process itself. Furthermore, \cite{reg11} showed that mass ratio distribution of visual stellar binaries down to $q\approx0.1$--0.2 depends only mildly, if at all, on stellar mass from low- to intermediate-mass stars. This suggests that the formation process of these binaries is the same across a broad range of stellar mass, enabling comparative studies while varying the mass of the primary star. While it was long believed that stars in a binary system are paired according to the initial (stellar) mass function, recent studies have shown that the mass ratio distribution is close to, and sometimes consistent with, a flat distribution \citep[e.g.,][]{reg13,Moe2017}. While the high-$q$ end of the distribution is generally well established, the frequency of ``extreme" systems (with $q \ll 1$) is much more difficult to sample due to observational challenges. At the same time, such highly unbalanced systems are rarely produced in numerical simulations as accretion on the initial ``seed binary" after the initial cloud fragmentation accounts for most of the stars' final mass \citep[e.g.,][]{bat12}. 

For solar-type stars, a dearth of companions in the brown-dwarf regime, dubbed the ``brown dwarf desert," has long been established through radial velocity and transit surveys \citep[e.g.,][]{gre06, sah11,Triaud2017, Feng2022}. Imaging surveys have confirmed that brown dwarf companions are also rare as visual companions to solar-type stars, although arguably not as much as at shorter orbital periods \citep{met09,bra14,cha15,Vigan2017, Bonavita2022b}. While early efforts were limited by the high contrast needed to detect non hydrogen-burning companions, possibly pointing to incompleteness, the most recent generation of instrumentation has become sufficiently sensitive to exclude this possibility, making the deficit of wide brown dwarf companions a reliable result \citep{Vigan2021}. 

The fact that the overall distribution of mass ratios presents a marked trough in the brown dwarf regime can be understood as the superimposition of two distinct regimes \citep{reg16}. On the one hand, the mass distribution of planetary-mass companions is heavily skewed towards low-mass objects \citep[e.g.,][]{cum08,Nielsen2019}, likely because of the limited amount of material available in protoplanetary disks. On the other hand, the mass ratio distribution of stellar visual binaries has a modest peak around $q\approx0.3$ and declines towards lower mass ratios over a broad range of primary masses, plummeting into the substellar regime \citep{Raghavan2010, Gullikson2016, Moe2017, ElBadry2019}. In other words, there appears to be a lower cutoff to the multiple star formation process. It may be that post-fragmentation accretion never occurs in a one-sided fashion, irrespective of the details of how the initial fragmentation developed. Alternatively, the fragmentation may occur in such a way that the initial seeds are already close to the stellar-substellar limit (instead of the often-quoted ``opacity limit" of a few Jupiter masses), so that relatively little subsequent accretion is needed to preclude the formation of a star-brown dwarf system. A broader characterization of ``extreme" binary systems thus has the potential to shed light on the early phases of (binary) star formation.

One of the key questions spurred by the existence of the brown dwarf desert is whether the limit of the stellar binary regime is a mass ratio limit, i.e., systems with $q\lesssim0.1$ are extremely difficult to form (``fixed mass ratio limit"), or a companion mass limit, i.e., systems with brown dwarf companions are intrinsically rare irrespective of the mass of the primary (``fixed companion mass limit"). Distinguishing these scenarios would provide important insight on the physics of star formation but is impossible while focusing exclusively on solar-type stars. Instead, samples of stars of different masses, for which the two limits would be cleanly separated, must be considered. The apparent lack of dependency of the mass ratio distribution with stellar mass supports the fixed mass ratio limit, but brown dwarf companions are generally treated as a unique category irrespective of the primary mass, as if the fixed companion mass was more relevant \citep[e.g.,][]{met09, Nielsen2019}. A dedicated survey is therefore necessary to shed empirical light on this question.

While the mass difference between a low-mass star and a putative brown dwarf companion is smaller than in the case of a solar-type target, the typically much older age of lower-mass stars lead to the companion being exceedingly faint. Thus, while some M dwarf--brown dwarf systems have been discovered \citep[e.g.,][]{nak95, Deacon2014, Han2022}, it is difficult to place them in a robust statistical context. For intermediate-mass stars, companions at the stellar/substellar limit correspond to mass ratios that are a factor of 2--5 lower than for solar-type stars. They therefore provide a more promising approach to determine whether the most extreme binary systems are set by a fixed mass ratio limit or a fixed companion mass limit. A number of systems containing an intermediate-mass primary and a brown dwarf companion at tens to hundreds of AU have been discovered using deep AO imaging \citep[e.g.,][]{low00,Huelamo2010,car13,der14,hin15,maw15,Wagner2020, Bonavita2022}, and their occurrence rate has been established through surveys using the latest generation of AO systems \citep{Nielsen2019,Vigan2021}. While these systems prove that substellar companions exist, they correspond to mass ratios of $q\lesssim0.05$, which correspond to the planetary mass regime for companion to solar-type stars. 

Determining the occurrence rate of (stellar) companions with $0.05 \lesssim q \lesssim 0.1$ for intermediate-mass stars is the key to distinguish between the fixed mass ratio limit and the fixed companion mass limit scenarios. Unfortunately, most previous multiplicity surveys of intermediate-mass stars focused on visual binaries have been limited by relatively small size samples \citep[$\lesssim40$ targets;][]{ehr10,vig12, Rameau2013}, by modest sensitivity to low-mass stellar companions at close separations \citep[$\lesssim200$\,AU;][]{kou05,rob07}, and/or significant confusion in regions like the Scorpius-Centaurus association \citep{sha02, kou07}. The largest survey for field intermediate-mass multiple systems, conducted by \cite{der14b}, was dominated by $\approx$1.6--2.1\,$M_\odot$ stars, offering limited leverage between the two scenarios discussed above. The same is true of the high-sensitivity planet-focused surveys of \cite{Nielsen2019} and \cite{Vigan2021}, which furthermore are biased against stellar binaries. Spectroscopic searches for cool companions to intermediate-mass stars also suffer from a rapid drop in sensitivity for $q\lesssim0.1$ \citep{Gullikson2016}. 

Given the limitations of previous surveys, we conducted a new dedicated survey for low-mass ($q\lesssim0.1$) stellar companions among 2--5\,$M_\odot$ in order to distinguish the fixed mass ratio and fixed companion mass limit scenarios. Specifically, we took advantage of the latest generation of high-resolution, high-contrast instruments installed at the Lick \citep[ShaneAO,][]{kup12,gav14} and Gemini \citep[Gemini Planet Imager, hereafter GPI,][]{Macintosh2014} observatories.
The paper is organized as follows: we describe our sample and observations in Section\,\ref{sec:sample_obs}, the data processing in Sections\,\ref{sec:shane_analysis} and \ref{sec:gpi_analysis}, the main results of our survey in Section\,\ref{sec:results} and their implications in Section\,\ref{sec:discus}. Finally, Section\,\ref{sec:concl} summarizes the main findings of our study.


\section{Sample and observations}
\label{sec:sample_obs}


\subsection{Sample selection}
\label{subsec:sample}

To target Main Sequence stars with masses $M_\star \gtrsim 2\,M_\odot$, our sample selection was focused on targets with early-A and B spectral type ($B-V \lesssim 0.1$). We drew targets from two complementary subsamples: field stars with {\it Hipparcos}-based distances $D\leq 100$\,pc (although subsequent {\it Gaia} data revised some distances above that limit), and members of a few open clusters. The cluster subsample, which was selected from the WEBDA open cluster database\footnote{http://www.univie.ac.at/webda/}, probes a potentially different star-forming environment, {generally characterized by different multiplicity properties due to dynamical interactions \citep[e.g.,][]{King2012, Marks2012}. As will be shown later, we do not find significant differences between the field and open cluster subsamples and will merge the two in a single overall sample. Crucially, open clusters probe younger stellar ages ($\lesssim$100\,Myr) which favors the inclusion of slightly more massive stars in the survey. The physical properties of the open clusters selected for this study are listed in Table\,\ref{tab:clusters}. We note that establishing membership to open clusters requires a holistic approach that combines photometry, spectroscopy and astrometry. This information is not uniformly available across the sample and it is likely that some of the objects studied here are actually not cluster members but instead unrelated field stars, as suggested by some of the individual distances being inconsistent with their supposedly parent cluster. With this caveat in mind, we leave individual cluster membership to further studies and consider the open clusters subsample as a coherent group.

\begin{table}
	\centering
	\caption{Properties of the open clusters included in this study. References for the distance and age of the clusters: 1) \citet{Bossini2019}; 2): \citet{Kovaleva2020}; 3) \citet{CantatGaudin2020}}
	\label{tab:clusters}
	\begin{tabular}{ccccccc} 
		\hline
		 & Cluster & $D$ & $t$ & Instrument & $N_{targets}$ & Ref. \\
		 &  & (pc) & (Myr) & & & \\
		\hline
        1 & Blanco\,1 & 237 & 95 & GPI & 2 & 1 \\
        2 & Collinder\,135 & 302 & 45 & GPI & 19 & 2 \\
        3 & Collinder\,359 & 552 & 37 & ShaneAO & 1 & 3 \\
        4 & IC\,2391 & 152 & 36 & GPI & 6 & 1 \\
        5 & IC\,4665 & 347 & 38 & ShaneAO & 3 & 1 \\
        6 & NGC\,2451\,A & 193 & 45 & GPI & 16 & 1 \\
        7 & Stephenson\,1 & 348 & 28 & ShaneAO & 10 & 1 \\
		\hline
	\end{tabular}
\end{table}

From both subsamples, we excluded targets that were known to be visual binaries with separations $\lesssim2$\arcsec\ and flux ratios $\lesssim2$\,mag, as these generally result in poorer AO correction. Furthermore, the presence of a bright companion severely restricts our ability to detect faint additional companions. In addition, we imposed magnitude limits to ensure high quality AO correction and to avoid saturation on the detector. For ShaneAO and GPI, this led to ranges of $3.5 \lesssim V \lesssim 9.5$ and $5.5 \lesssim V \lesssim 9$, respectively. Finally, a few stars that passed these criteria were rejected a posteriori due to inconsistent distance estimates; these objects are listed in Appendix\,\ref{sec:rejected}. Their images were used in the PSF subtraction process but they were not included in any statistical analysis presented here. The complete sample contains 169 targets, 57 of which are located in open clusters; 126 and 43 targets were observed with ShaneAO and GPI, respectively. The median distance in the overall sample is 105\,pc and drops to 87\,pc for the field stars subsample. Finally, the sample covers the B3--A5 spectral type range and 24 systems are known as spectroscopic binaries or triple systems \citep{Pourbaix2004}, although we note that this census is most likely incomplete.

\begin{figure*}
\includegraphics[width=0.45\textwidth]{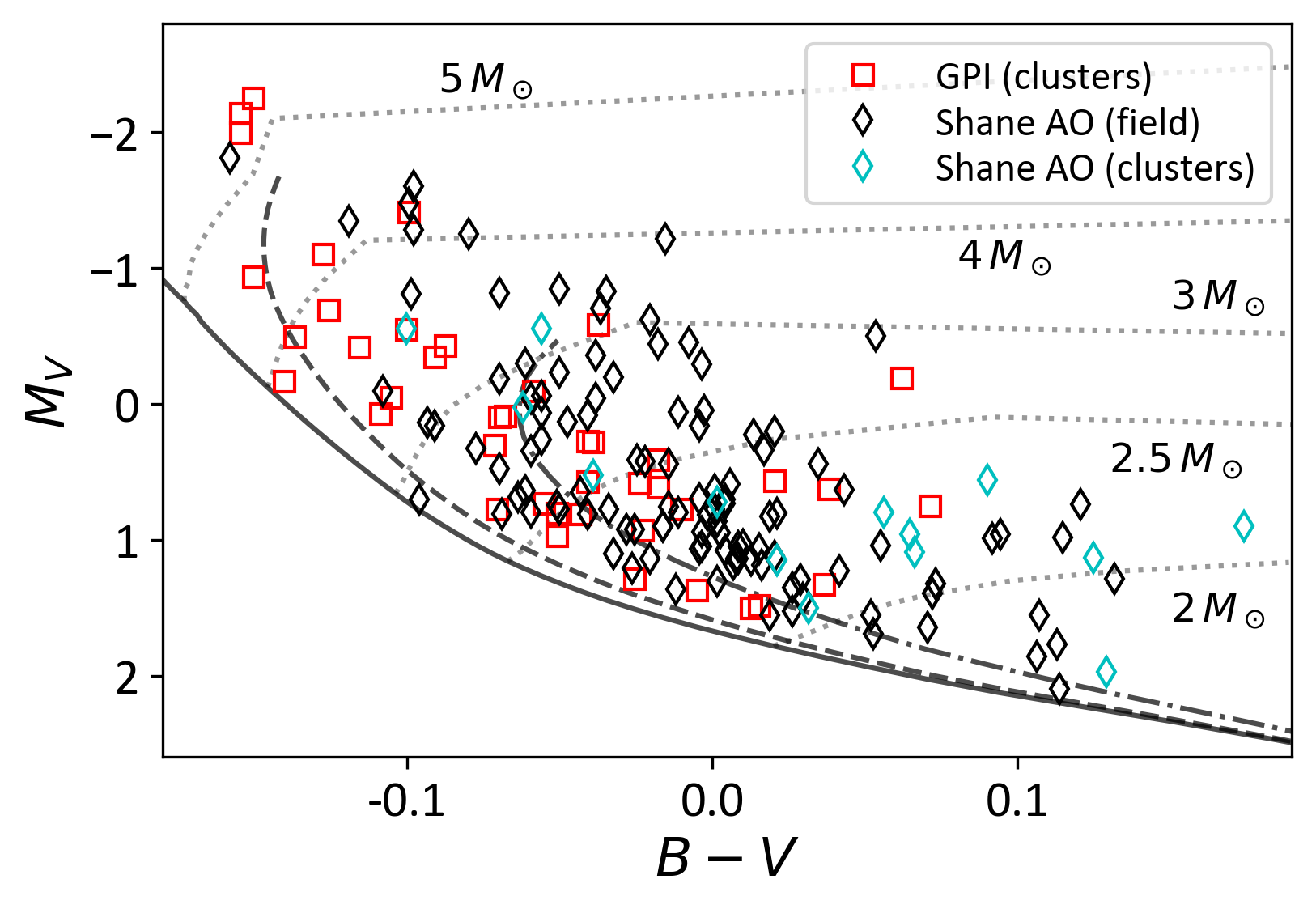}\hspace*{1cm}
\includegraphics[width=0.45\textwidth]{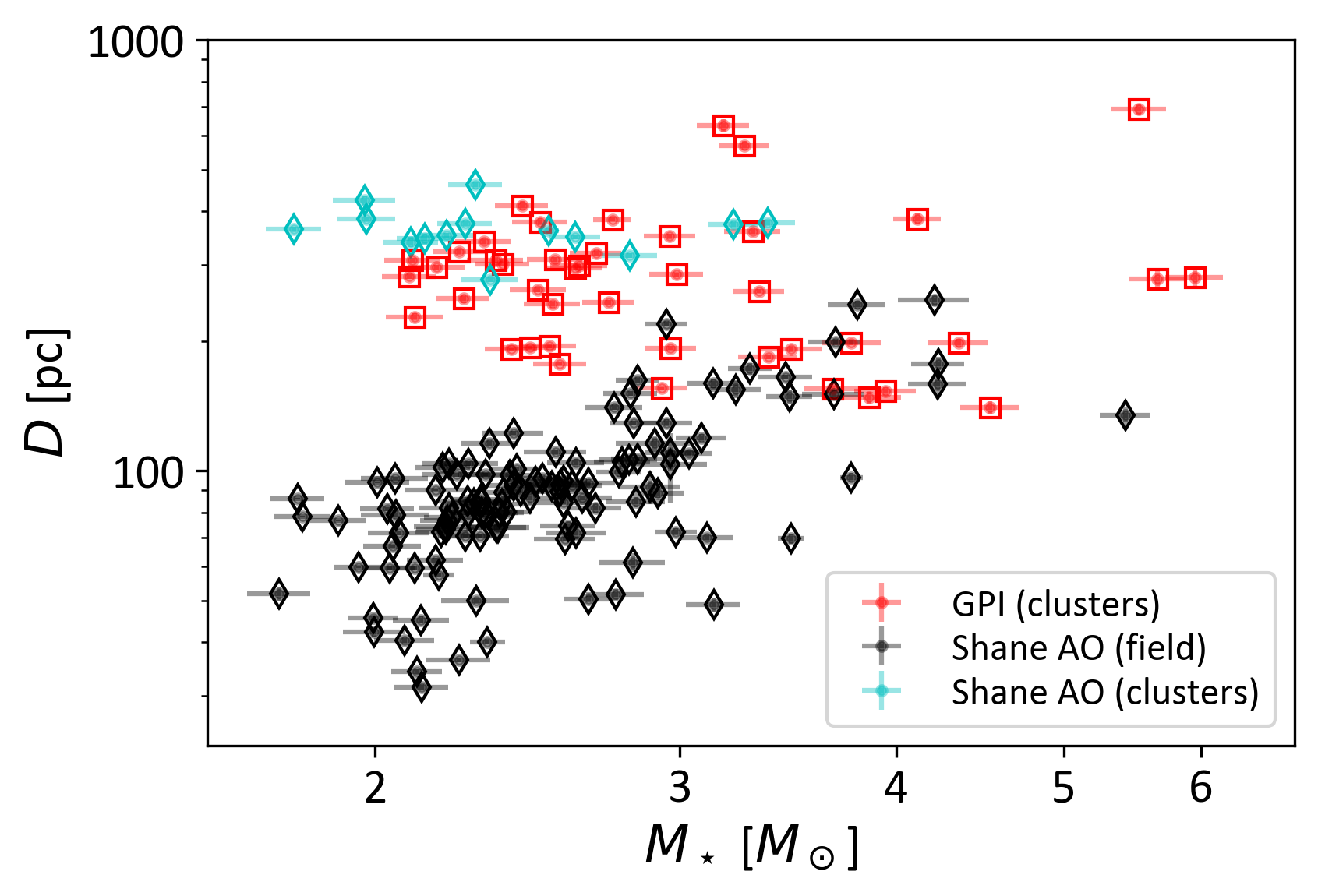}
\caption{{\it Left:} Color-magnitude diagram for all stars observed in this survey. Colored symbols disentangle field stars from members of open clusters. The solid, dashed and dot-dashed curves represent the 30, 100 and 300\,Myr solar metallicity isochrones based on the MIST evolutionary models, respectively, whereas dotted curves show the evolutionary tracks for stars of masses ranging from 2 to 5\,$M_\odot$. {\it Right:} Mass and distance distributions for our observed sample. Errorbars represent the 68-percentile confidence interval based on our MCMC fit to the targets' photometry and parallax. \label{fig:sample}}
\end{figure*}

We used the same approach as in \citet{Nielsen2019} to estimate the age and mass of each star within the sample (see their Section 2.1.1). We opted to use both {\it Gaia} DR2 photometry ($G_B$, $G$, $G_R$; Gaia Collaboration et al. 2018) and Tycho2 ($B_T$, $V_T$; Hog et al. 2000) photometry, as these were available for all stars within the sample. Unlike \citet{Nielsen2019}, we did not use a prior on the age if the star was a known member of an open cluster as membership cannot be determined with parallax alone, nor did we use one on the projected rotational velocity of the star. We allowed metallicity to vary in the range $-2 \leq [\mathrm{Fe/H}] \leq 0.5$ but found that the vast majority of target stars were close to solar metallicity, with the mean and standard deviation of all best-fit mettalicity being -0.03 and 0.13\,dex, respectively. Using an MCMC approach to the fit, we derive posterior distributions for the mass of each target, which are listed in Table\,\ref{tab:sample_shane} and \ref{tab:sample_gpi} and are shown in Fig.\,\ref{fig:sample}. 

The inferred masses range from $\approx1.75\,M_\odot$ to $\approx4.5\,M_\odot$ with four outliers above $5\,M_\odot$ and a median mass of $\approx2.5\,M_\odot$. As a general rule, the MCMC posteriors show an anticorrelation between stellar age and mass, as expected from stellar evolution models. In the most extreme cases, this effect can lead to uncertainties of up to $\approx15$\% on the stellar mass when the object is near, or possibly past, the end of their Main Sequence lifetime. Furthermore, our procedure assumes no foreground extinction, which is likely incorrect for the most distant targets. We have performed stellar parameters fits including extinction for a few stars and found that the effect is to increase the stellar masses by 5-10\% for the most extincted objects in our sample, albeit with a similar increase in random errors due to unaccounted for systematic errors in absolute photometric calibration. Similarly, taking into account of neglecting the presence of an unresolved (spectroscopic) companion induces errors on the primary mass at the 5\% level. Since these effects are propagated to the companion masses we infer in Section\,\ref{subsec:comp_mass}, we conclude that extinction and unresolved multiplicity have a negligible effect on the mass ratios (uncertainty of 0.01 or lower) derived in this study and decided to use the extinction-free, single star, stellar parameters estimates for practical purposes. More detailed analysis, including spectroscopic information, would be needed to obtain more precise estimates. Given the small number of targets affected, the limited amount of photometric data at hand, and to ensure consistency in the analysis, we retain the MCMC-based mass estimates for all objects.

\begin{table*}
	\centering
	\caption{Properties of the targets observed with ShaneAO. The mass, distance and age of each targets (third through fifth column) are derived from our analysis of photometric and parallax measurements (see Section\,\ref{subsec:sample}). The eighth column indicates the number of useful exposures in the complete observing sequence. A numbered note in the last column indicates that the target is a (candidate) member of the corresponding open cluster in Table\,\ref{tab:clusters}; otherwise it is a field star. Furthermore, "SB" indicates systems listed in the 9th Catalog of Spectroscopic Binaries \citep{Pourbaix2004}. 
	Only the first few lines of the table are shown here; the entire table is available electronically.}
	\label{tab:sample_shane}
	\begin{tabular}{cccccccccc} 
		\hline
Target & Sp.T. & $M_\star$ & $D_{MCMC}$ & $t_{MCMC}$ & Obs. Date & Filter & $t_{exp}$ & $N_{exp}$ & Note \\
 & & ($M_\odot$) & (pc) & (Myr) & (UT) & & (s) & & \\
		\hline
BD\,+36\,3314  & A0 &  2.25 $^{+0.09}_{-0.08}$ &  374.3 $^{+6.3}_{-5.9}$ &  616 $^{+55}_{-62}$ &  2015-07-04  &  $K_s$  & 5$\times$2 &  46 / 100  & 7 \\ 
BD\,+36\,3317  & A0 &  2.52 $^{+0.03}_{-0.04}$ &  360.3 $^{+5.1}_{-5.0}$ &  181 $^{+47}_{-33}$ &  2015-07-04  &  $K_s$  & 10 &  93 / 100  & 7 \\ 
BD\,+36\,3274  & A0 &  1.98 $^{+0.08}_{-0.07}$ &  383.6 $^{+5.4}_{-5.3}$ &  522 $^{+111}_{-109}$ &  2015-07-04  &  $K_s$  & 10 &  99 / 100  & 7 \\ 
BD\,+36\,3313  & B9--A0 &  1.79 $^{+0.07}_{-0.06}$ &  363.0 $^{+4.7}_{-4.7}$ &  470 $^{+185}_{-182}$ &  2015-07-04  &  $K_s$  & 10 &  99 / 100  & 7 \\ 
HD\,6456  & A0IV-Vnn &  2.32 $^{+0.11}_{-0.10}$ &  79.9 $^{+4.9}_{-4.2}$ &  481 $^{+64}_{-67}$ &  2016-10-26  &  Br$_\gamma$  & 5 &  53 / 100  & \\
HD\,13294  & B9V &  2.75 $^{+0.11}_{-0.10}$ &  140.3 $^{+2.0}_{-2.0}$ &  292 $^{+40}_{-41}$ &  2016-09-19  &  Br$_\gamma$  & 10 &  98 / 100  & \\ 
HD\,14212  & A0V &  2.37 $^{+0.08}_{-0.07}$ &  83.6 $^{+1.2}_{-1.2}$ &  477 $^{+50}_{-52}$ &  2016-09-19  &  Br$_\gamma$  & 3 &  99 / 100  & \\ 
HD\,62832  & A1Vn &  2.50 $^{+0.09}_{-0.08}$ &  95.9 $^{+1.8}_{-1.9}$ &  502 $^{+34}_{-42}$ &  2016-03-27  &  Br$_\gamma$  & 1.5 &  123 / 125  & \\ 
HD\,64648  & A0Vs &  3.29 $^{+0.09}_{-0.10}$ &  172.3 $^{+4.8}_{-5.0}$ &  292 $^{+21}_{-18}$ &  2016-03-26  &  Br$_\gamma$  & 1.5 &  135 / 136  & \\ 
HD\,66664  & A1V &  2.17 $^{+0.09}_{-0.08}$ &  61.9 $^{+0.9}_{-0.9}$ &  441 $^{+83}_{-90}$ &  2016-03-27  &  Br$_\gamma$  & 1.5 &  119 / 125  & \\ 		\hline
	\end{tabular}
\end{table*}

\begin{table*}
	\centering
	\caption{Properties of the targets observed with GPI. Columns and notes have the same meaning as in Table\,\ref{tab:sample_shane}, except for the eighth column which indicates the amount of field rotation recorded during the data acquisition for each system.
	Only the first few lines of the table are shown here; the entire table is available electronically.}
	\label{tab:sample_gpi}
	\begin{tabular}{ccccccccc} 
		\hline
Target & Sp.T. & $M_\star$ & $D_{MCMC}$ & $t_{MCMC}$ & Obs. Date & $N_{exp}$ & Field Rot. & Note \\
 & & ($M_\odot$) & (pc) & (Myr) & (UT) & & (\degr) & \\
		\hline
HD 55742  &  A0  &  2.74 $^{+0.06}_{-0.08 }$ &  382.7 $^{+4.3}_{-4.3 }$ &  466 $^{+34}_{-26 }$ &  2016-01-22  &  8  &  2.0  & 2 \\ 
HD 56283  &  B9V  &  2.09 $^{+0.08}_{-0.07 }$ &  282.6 $^{+3.8}_{-3.7 }$ &  254 $^{+117}_{-106 }$ &  2016-01-21  &  2  &  2.7  & 2 \\ 
HD 56284  &  B8  &  5.53 $^{+0.17}_{-0.23 }$ &  689.6 $^{+21.8}_{-21.3 }$ &  72 $^{+10}_{-5 }$ &  2016-01-22  &  7  &  3.7  & 2 \\ 
HD 56346  &  A0V  &  2.10 $^{+0.08}_{-0.07 }$ &  307.5 $^{+4.7}_{-4.5 }$ &  408 $^{+101}_{-103 }$ &  2015-12-26  &  8  &  18.3  & 2 \\ 
HD 56376  &  B6V  &  2.99 $^{+0.10}_{-0.11 }$ &  285.2 $^{+3.4}_{-3.1 }$ &  235 $^{+34}_{-31 }$ &  2016-01-22  &  6  &  0.9  & 2 \\ 
HD 56474  &  B9V  &  2.43 $^{+0.09}_{-0.08 }$ &  412.8 $^{+10.0}_{-9.0 }$ &  471 $^{+46}_{-52 }$ &  2015-12-25  &  6  &  18.3  & 2 \\ 
HD 56557  &  B9V  &  2.17 $^{+0.08}_{-0.08 }$ &  296.0 $^{+4.4}_{-4.4 }$ &  273 $^{+102}_{-101 }$ &  2016-01-19  &  7  &  2.8  & 2 \\ 
HD 56854  &  A0V  &  2.35 $^{+0.09}_{-0.08 }$ &  307.5 $^{+5.5}_{-5.1 }$ &  366 $^{+67}_{-69 }$ &  2016-01-23  &  7  &  12.1  & 2 \\ 
HD 56856  &  B9.5V  &  2.37 $^{+0.08}_{-0.09 }$ &  301.9 $^{+4.1}_{-4.1 }$ &  462 $^{+57}_{-55 }$ &  2015-12-25  &  6  &  1.5  & 2 \\ 
HD 56932  &  B7/B8V  &  2.61 $^{+0.10}_{-0.10 }$ &  295.9 $^{+3.9}_{-3.8 }$ &  298 $^{+49}_{-54 }$ &  2015-12-25  &  6  &  2.3  & 2 \\
\hline
	\end{tabular}
\end{table*}


\subsection{Observations and initial data reduction}

\subsubsection{ShaneAO observations}

Observations were conducted at the Shane 3m telescope with the ShaneAO AO system feeding the ShARCS near-infrared instrument \citep{mcg14} over the course of eight observing runs from June 2015 to May 2018. We conducted the survey exclusively at 2\,$\mu$m as this ensures 1) the uniformity of datasets and processing for all our targets, 2) a high quality AO correction, and 3) the possibility of building a large library of single stars that can serve as point spread function (PSF) references. The latter point is dictated by the fact that the instrumental setup does not allow for field rotation, preventing the use of angular differential imaging (ADI; \citealp{Marois2006}), a standard approach to achieve high contrast with AO imaging. Instead, we resort to reference differential imaging (RDI), i.e., observations of single stars will be used to generate the PSF corresponding to a given target (see Sect\,\ref{subsec:shane_psfsub}).

Depending on the target's brightness, we used the broad-band \ks ($\lambda_c=2.15\mu$m, $\Delta\lambda=0.32\mu$m) or narrow-band \brg ($\lambda_c=2.167\mu$m, $\Delta\lambda=0.020\mu$m) filters, in combination with short exposure times, to avoid saturation of the core of the PSF. This is motivated by our goal to achieve high contrast as close to the central star as possible. An alternative method consists in using coronagraphy and longer individual exposures, but this increases the inner working angle at which faint companions could be found and introduces additional overheads. To achieve sensitivity to faint companions, we employed an observing sequence that accumulates 75 to 125 images per target, dithered at 5 positions on the detector to reduce the effect of bad pixels and estimate the sky background. In some cases, several images were coadded before the resulting image was saved on disk. A detailed log of observations is presented in Table\,\ref{tab:sample_shane}.

All data were reduced following standard procedures for near-infrared datasets, with sky subtraction based on the median of all frames in a sequence and flat-fielding using twilight sky observations. After discarding the occasional frames that have a saturated core, all images in a given sequence were then aligned based on the location of the target's light centroid. Frame selection was then performed, with the requirement that the peak pixel value be at least 50\% of the highest in the entire stack. This selection is intended to discard images taken during periods of markedly inferior AO performance, which negatively affect the resulting contrast. The number of frames that passed this image quality selection is indicated in Table\,\ref{tab:sample_shane}; the median (mean) frame acceptance fraction across all useful datasets is 97\% (88\%). A deep image of each system was then produced by median-collapsing each datacube. The stacked \brg images are characterized by a high quality AO correction, with up to 12 (partial) Airy rings in the best cases. The broadband \ks images are also of excellent quality, although the PSF radial profile is slightly smeared due to the much wider bandpass.

Since the astrometric characteristics of ShARCS (plate scale and absolute orientation, as well as their run-to-run stability) were unknown when we started this project, we obtained observations of wide binaries to serve as astrometric calibrators. The calibration binaries were selected from the Washington Double Star (WDS) orbit catalog\footnote{http://www.astro.gsu.edu/wds/}. The detailed log of these observations and their analysis is described in Appendix\,\ref{sec:shaneao_calib}. From these binaries, we determined the plate scale of ShARCS to be 32.63$\pm$0.12\,mas/pix and established its absolute orientation to within 0\fdg09.

\subsubsection{GPI observations}

GPI integral field spectroscopy observations were obtained between December 2015 and January 2016 as part of program GS-2015B-Q-37 with a standard procedure for all objects. We selected the $H$ band filter and its associated apodized Lyot coronagraph as this combination provides the best compromise between AO performance and angular resolution. The full field of view of each dataset is 2\farcs8 on the side with a lenslet spacing of 0\farcs1416 \citep{DeRosa2020}, and the spectral coverage of the integral field spectrograph ranges from $\approx$1.5$\,\mu$m to $\approx$1.8$\,\mu$m with a spectral resolution of $\Delta\lambda / \lambda \approx 40$--50. The integration time was set to 60\,s to avoid saturation near the edge of the coronagraphic mask. A few (5--8) images were saved in succession for each object except when a bright companion was readily apparent in the first frame; in those cases, the sequence was aborted and a single exposure was recorded. Because GPI is designed to enable ADI, the field rotates as the telescope tracks the target in the sky. However, due to the snapshot nature of our observing sequences, the amount of field rotation is limited (median of $\approx4$\degr, see Table\,\ref{tab:sample_gpi}). Observations of an argon lamp were taken during the target acquisition to measure the instrument flexure induced by the changing gravity vector as the telescope changes position. Standard dark and wavelength calibration frames were taken during the daytime as a part of the observatory's calibration plan. 

The data reduction was performed using the GPI Data Reduction Pipeline (DRP; \citealp{Perrin2014, Wang2018}) v1.3.0. In short, the individual frames are dark subtracted and bad pixels are identified and interpolated over. The microspectra within each image are then extracted using the location of the argon spectra to produce 3-dimensional $(x,y,\lambda)$ datacubes, and an additional bad pixel correction is performed before optical distortion is corrected based on a static distortion map applied to each slice within the data cube. Finally, the location and brightness of the (occulted) star is determined in each slice of the datacube based on the location of so-called satellite spots generated by a wire diffraction grid within the pupil plane \citep{Sivaramakrishnan2006, Marois2006b}. These satellite spots have been measured to be 9.40$\pm$0.02\,mag fainter than the star itself \citep{DeRosa2020b}. We note that the absolute orientation of the GPI instrument is slightly inaccurate with the version of the pipeline used here and we subsequently applied the correction appropriate for the epoch of observation as determined by \citet{DeRosa2020}.


\section{ShaneAO Data Analysis}
\label{sec:shane_analysis}


\subsection{Visual inspection}

Visual inspection of the deep ShaneAO images revealed the presence of 12 companions with separations ranging from $\approx$0\farcs35 to $\approx$2\arcsec. We evaluate the relative astrometry of the binary by computing the centroid location of the primary star in the direct image and that of the companion after subtracting a radial profile approximating the PSF of the primary. We conservatively estimate astrometric uncertainty of 0.1\,pixel, or 0\farcs003. The binary flux ratio is measured by applying aperture photometry with a 3-pixel radius aperture, with an uncertainty derived from the scatter from apertures placed at the same distance from the primary star but at different position angles. Furthermore, a ``floor" uncertainty of 0.1\,mag, consistent with the scatter of the differences between individual epochs of repeat observations, is applied where necessary. We assume that the magnitude difference measured in either the narrow Br$_\gamma$ and the broad $K_s$ filter are both good estimates of the standard $K$ flux ratio. All results are listed in the top part of Table\,\ref{tab:shane_bins}. The brighter of these companions, specifically those with $\Delta K \lesssim 4$\,mag, were all previously known, as tabulated in the Washington Double Star Catalog \citep[WDS,][]{Mason2001}. We note that that catalog includes physical as well as optical companions and thus inclusion in the WDS is insufficient to infer that a system is bound. 

\begin{table}
	\centering
	\caption{Relative astrometry and photometry of the binary systems detected in ShaneAO images. Companion previously listed in the WDS catalog are indicated with a $\dagger$ symbol.}
	\label{tab:shane_bins}
	\begin{tabular}{ccccc} 
		\hline
		Target & Obs. Date & Sep. & PA & $\Delta K$ \\
		 & (UT) & (\arcsec) & (\degr) & (mag) \\
		\hline
		\multicolumn{5}{c}{Companions identified in direct images}\\
		\hline
HD\,148112$^\dagger$ & 2017-05-16 & 0.849$\pm$0.004 & 302.61$\pm$0.22 & 3.37$\pm$0.11 \\
 & 2018-05-27 & 0.857$\pm$0.004 & 303.72$\pm$0.22 & 3.42$\pm$0.11 \\
HD\,161734 & 2015-06-06 & 1.888$\pm$0.008 & 117.87$\pm$0.13 & 6.17$\pm$0.10 \\
 & 2018-05-27 & 1.880$\pm$0.008 & 118.06$\pm$0.13 & 6.10$\pm$0.10 \\
HD\,204770$^\dagger$ & 2016-09-20 & 0.603$\pm$0.004 & 321.17$\pm$0.30 & 3.71$\pm$0.22 \\
 & 2017-10-04 & 0.608$\pm$0.004 & 321.13$\pm$0.30 & 3.60$\pm$0.22 \\
HD\,221253$^\dagger$ & 2016-09-19 & 0.717$\pm$0.004 & 0.98$\pm$0.26 & 2.88$\pm$0.16 \\
 & 2017-10-04 & 0.715$\pm$0.004 & 1.05$\pm$0.026 & 3.02$\pm$0.16 \\
HIP\,3544\,AB$^\dagger$ & 2016-09-19 & 1.543$\pm$0.006 & 172.90$\pm$0.14 & 3.09$\pm$0.10 \\
 & 2017-10-04 & 1.536$\pm$0.004 & 172.85$\pm$0.14 & 3.13$\pm$0.10 \\
HIP\,5310$^\dagger$ & 2017-10-05 & 0.557$\pm$0.004 & 214.23$\pm$0.32 & 4.04$\pm$0.24 \\
HIP\,13775 & 2015-12-01 & 1.061$\pm$0.005 & 45.58$\pm$0.19 & 4.55$\pm$0.10 \\
 & 2017-10-04 & 1.063$\pm$0.005 & 45.670.019 & 4.67$\pm$0.10 \\
HIP\,20648$^\dagger$ & 2017-10-05 & 1.802$\pm$0.007 & 342.91$\pm$0.13 & 2.36$\pm$0.10 \\
HIP\,22842 & 2015-12-01 & 0.895$\pm$0.004 & 339.37$\pm$0.21 & 4.66$\pm$0.10 \\
 & 2017-10-04 & 0.914$\pm$0.005 & 340.18$\pm$0.21 & 4.92$\pm$0.10 \\
HIP\,107253$^\dagger$ & 2015-07-03 & 1.899$\pm$0.008 & 142.23$\pm$0.13 & 2.92$\pm$0.10 \\
 & 2015-11-29 & 1.906$\pm$0.008 & 142.14$\pm$0.13 & 2.90$\pm$0.10 \\
 & 2017-10-05 & 1.914$\pm$0.008 & 142.28$\pm$0.13 & 2.86$\pm$0.10 \\
HIP\,109745 & 2016-09-20 & 1.533$\pm$0.006 & 16.60$\pm$0.14 & 3.98$\pm$0.10 \\
 & 2017-10-04 & 1.529$\pm$0.006 & 15.90$\pm$0.14 & 3.91$\pm$0.10 \\
HIP\,109831$^\dagger$ & 2017-10-04 & 1.402$\pm$0.006 & 298.42$\pm$0.15 & 4.51$\pm$0.10 \\
 & 2018-05-27 & 1.403$\pm$0.006 & 298.73$\pm$0.15 & 4.36$\pm$0.10 \\
        \hline
		\multicolumn{5}{c}{Companions identified in PSF-subtracted images}\\
		\hline
HD\,130109 & 2017-05-16 & 0.603$\pm$0.005  & 97.49$\pm$0.22  & 4.62$\pm$0.22 \\
HD\,152614 & 2016-03-27 & 0.355$\pm$0.009  & 73.06$\pm$0.51  & 4.06$\pm$0.21 \\
 & 2018-05-27 & 0.365$\pm$0.009  & 75.77$\pm$0.51  & 4.30$\pm$0.22 \\
HIP\,3544\,AaAb & 2016-09-19 & 0.646$\pm$0.005  & 245.95$\pm$0.22  & 4.64$\pm$0.20 \\
 & 2017-10-04 & 0.667$\pm$0.005  & 246.27$\pm$0.22  & 4.45$\pm$0.20 \\
HIP\,92312 & 2017-05-16 & 0.388$\pm$0.009  & 10.39$\pm$0.51  & 4.82$\pm$0.23 \\
HIP\,116611$^\dagger$ & 2015-12-02 & 0.948$\pm$0.005 &  169.57$\pm$0.11  & 6.16$\pm$0.20 \\
 & 2017-10-05 & 0.941$\pm$0.005 & 168.75$\pm$0.11  & 5.94$\pm$0.20 \\
        \hline
	\end{tabular}
\end{table}

\begin{figure*}
\includegraphics[width=0.32\textwidth]{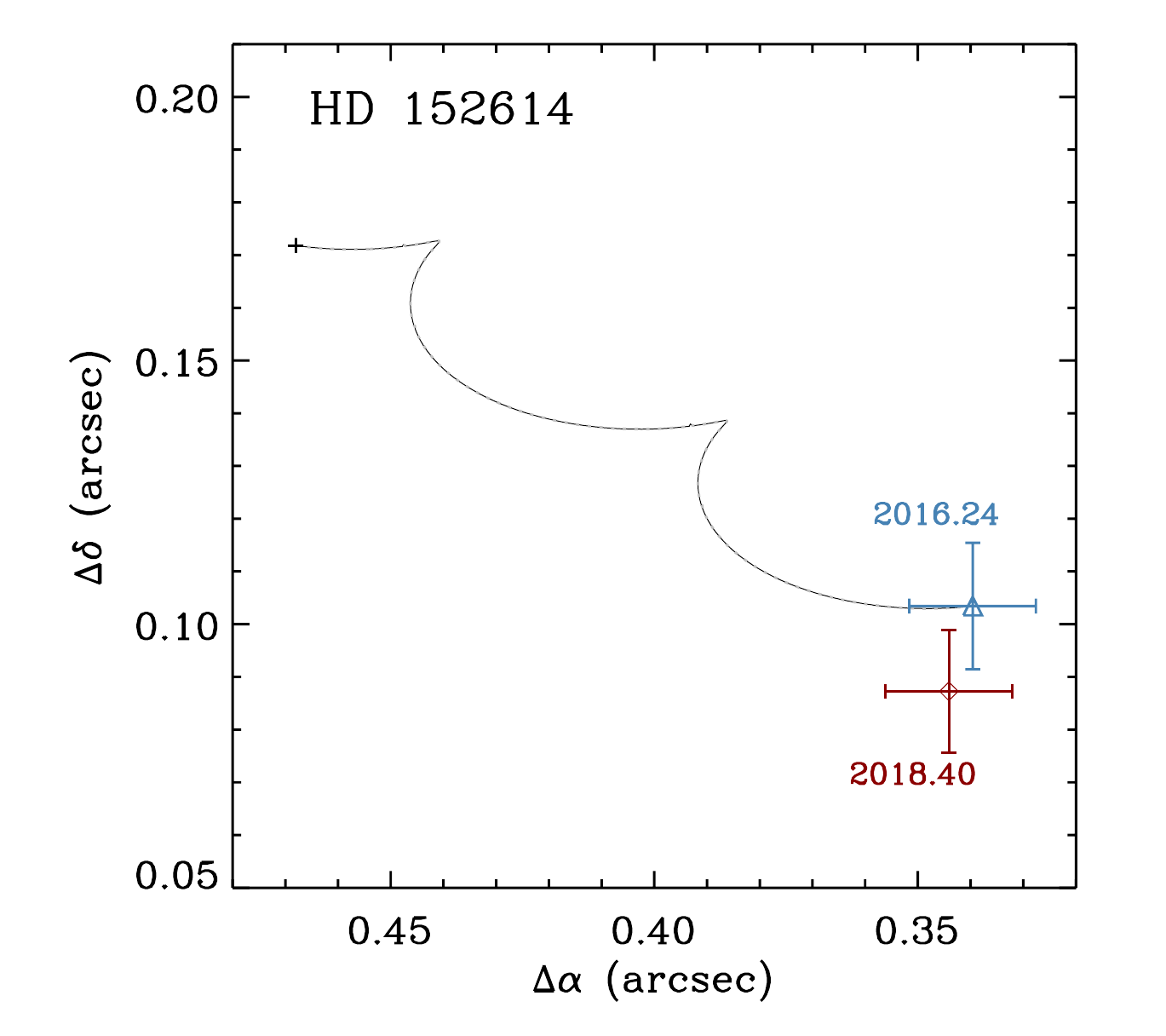}
\includegraphics[width=0.32\textwidth]{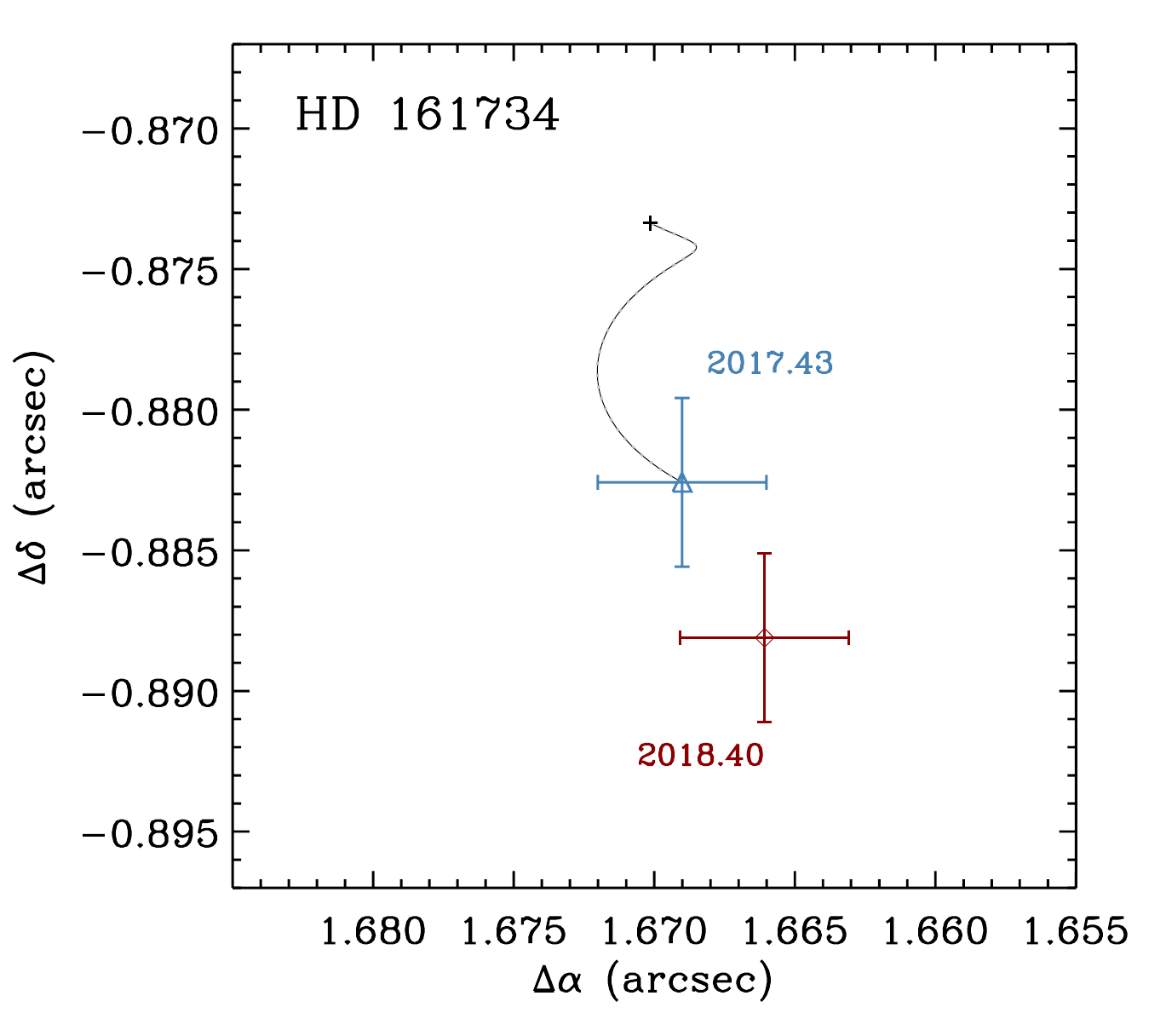}
\includegraphics[width=0.32\textwidth]{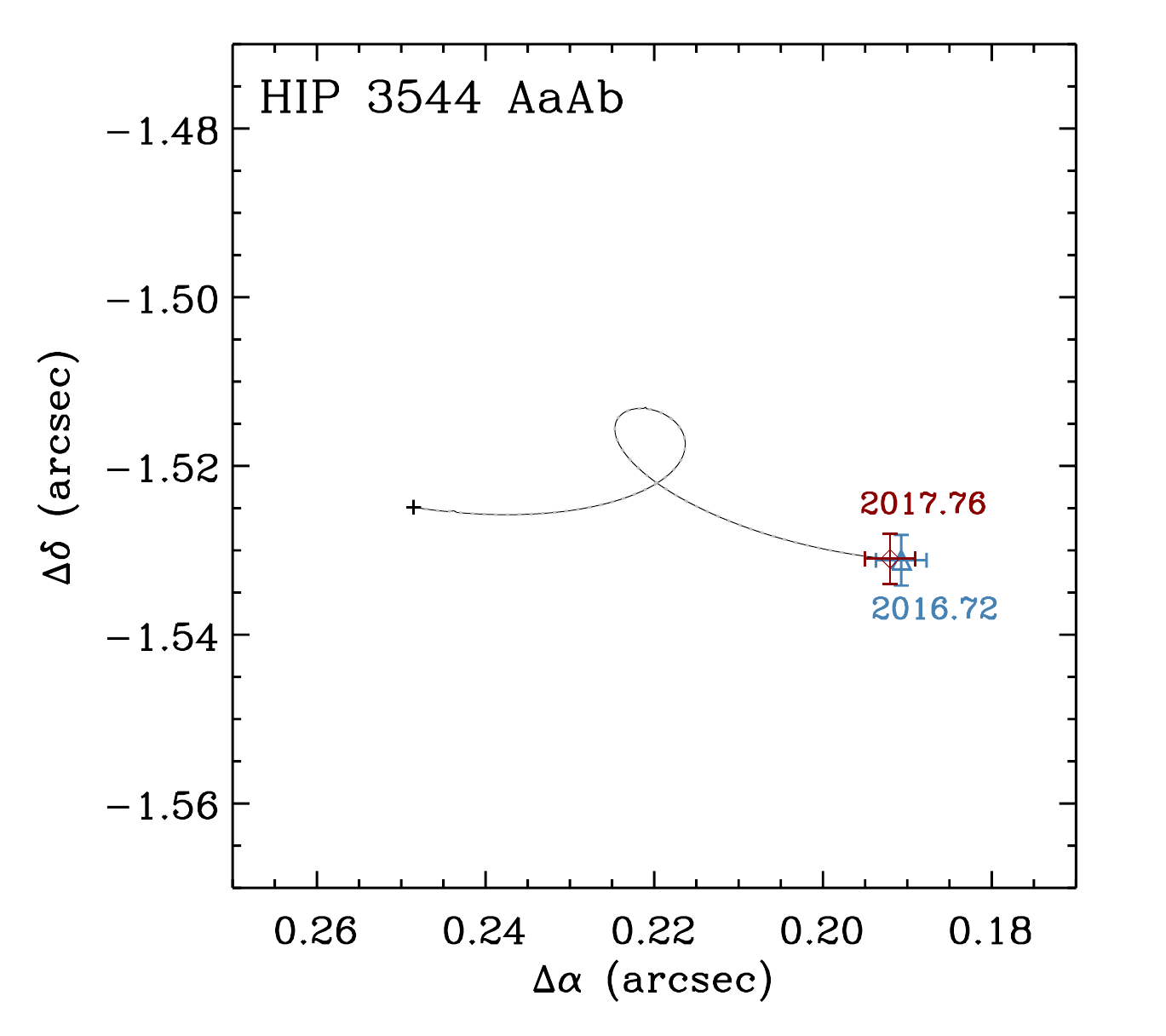}\\
\includegraphics[width=0.32\textwidth]{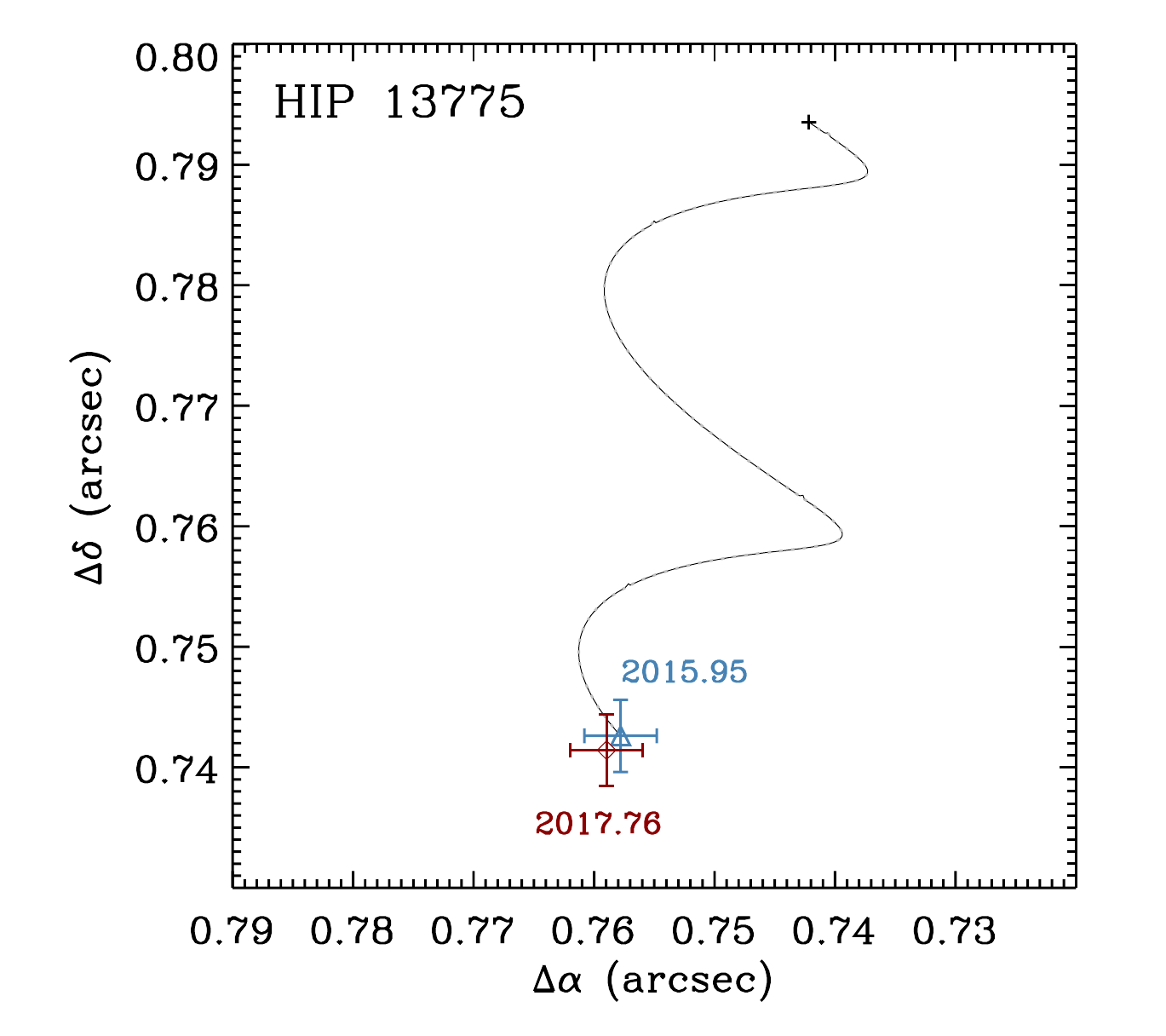}
\includegraphics[width=0.32\textwidth]{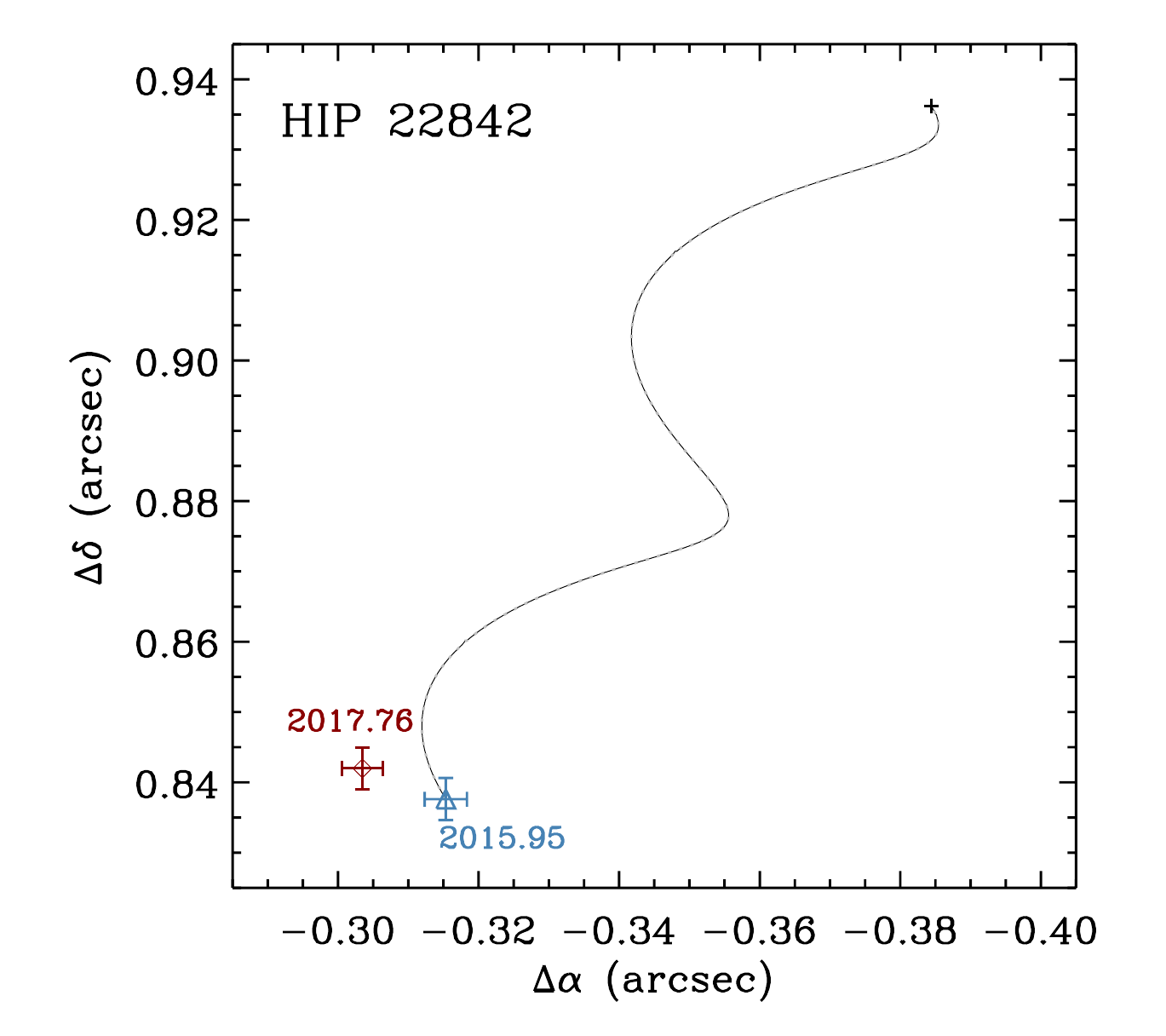}
\includegraphics[width=0.32\textwidth]{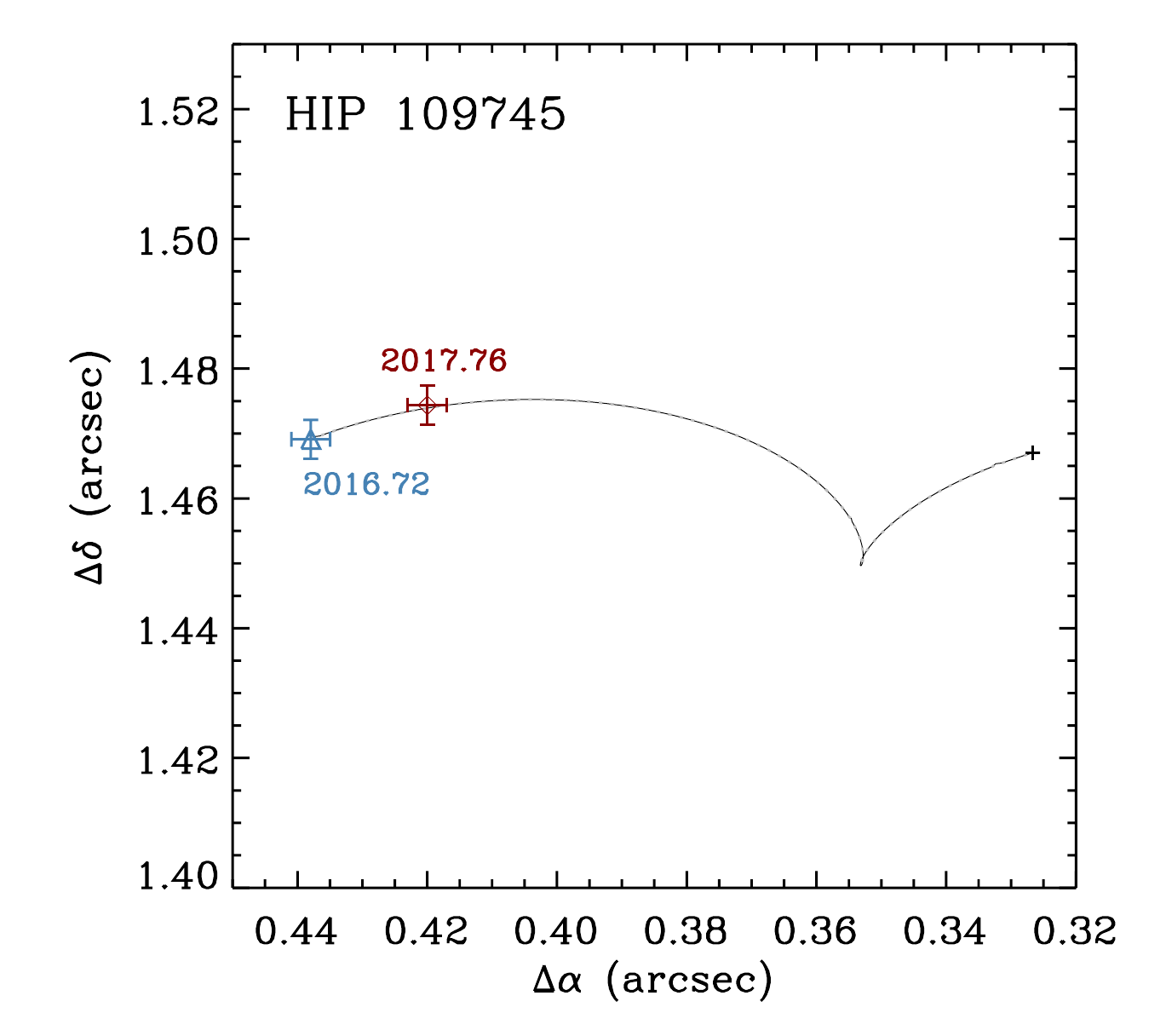}\\
\caption{Relative astrometry of the companions discovered during ShaneAO observations. Light blue triangles and dark red diamonds represent the first and second epoch of observation, respectively. The black curve represents the expected motion of the companion over the same period if it weer a static background stars. The integrated uncertainties on the target's parallax and proper motion are rendered as the black cross at the end of each track. \label{fig:common_motion}}
\end{figure*}

Since most of those companions were detected in real time at the telescope, we obtained a second epoch during subsequent observing runs, 1--2 years after the first observation, to test for common proper motion. Given the relatively large proper motion of our targets, $\gtrsim$1\,pix/yr, we find that all companions are inconsistent with background sources and, in some cases, detect non-negligible orbital motion. Fig.\,\ref{fig:common_motion} illustrates the apparent motion of the newly discovered companions in comparison to expectation for a non-moving background star. In all cases, the second epoch of observations is at least 4$\sigma$ discrepant with the background star assumption. This analysis confirms that all of these are physically associated to the survey targets.

It is impossible to conclusively assess whether a system is physical on the basis of a single detection. However, physically bound companions can induce a significant reflex motion on their primary star, which can result in the system being an astrometrically accelerating one. We search all systems for significant motion in catalogs comparing their {\it Gaia} and {\it Hipparcos} proper motion \citep[e.g.,][]{Brandt2021}. We found that both HIP\,5310 and HIP\,20648 have a significant ($\geq 5\,\sigma$) acceleration in the \cite{Kervella2022} catalog, providing support to the hypothesis that these systems are physical. For context, 7 of the 10 systems with two epochs of observations display a detectable ($\geq3\,\sigma$) acceleration in the same catalog, even though all of them show common proper motion. Despite conclusive evidence to support, we assume that the other companions are also physically bound given that they span similar ranges of separations and flux ratios.

Finally, we noticed that HD\,6456 appeared consistently elongated through the entire observing sequence obtained on 2016-10-26, indicating that it is a close binary (see Fig.\,\ref{fig:close_bins}). We then performed PSF fitting using all single stars observed in the same run and with the same filter (a total of 10 PSF stars) and fitting for the relative position and flux ratios of two components. This led to a binary separation, PA, and $K$ magnitude difference of 0\farcs133$\pm$0\farcs007, 308$\pm$5\degr, and 0.43$\pm$0.10\,mag, respectively. This is fully consistent with the known orbit of the system \citep{Horch2010}. We further find tentative evidence that HD\,173524 may be a very close binary at position angle $\sim$180\degr, but the PSF method outlined above failed to provide consistent results when using different PSF stars. The system was detected as an 0\farcs2 binary with a position angle of 135\degr\ in 1987 \cite{Isobe1990}, but it appears to be too challenging a system to confidently detect with ShaneAO at the epoch of our observations.

\begin{figure}
\centering
\includegraphics[width=0.8\columnwidth]{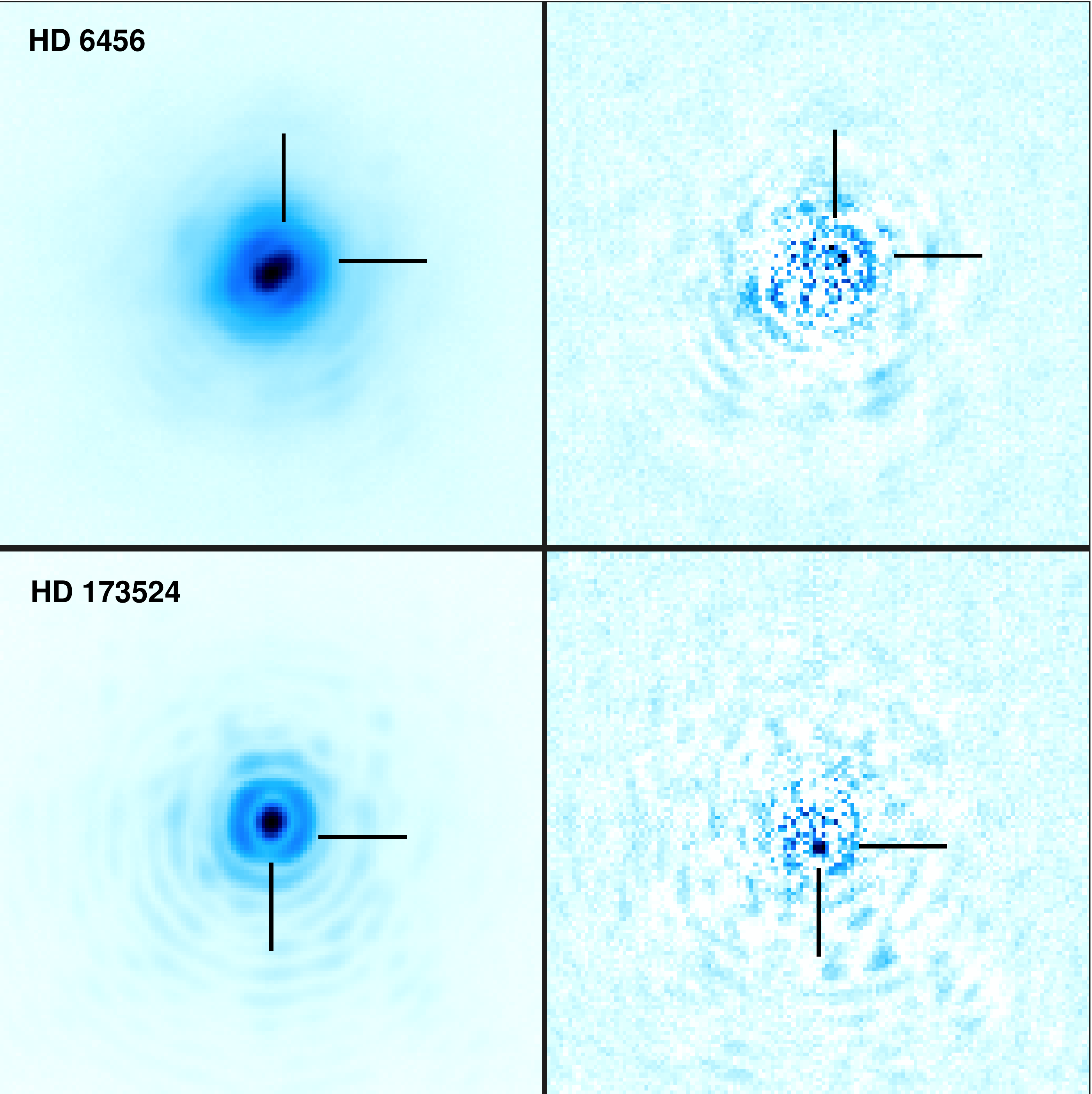}\\
\caption{Close binaries detected based on the observed elongation of the core of the PSF during ShaneAO observations. The left and right columns represent the direct and LOCI-processed final images of each system, displayed with a square root and linear stretch, respectively. Core elongation around PA $\approx$310\degr is clear for HD\,6456 whereas the modest North-South elongation of HD\,173524 is confirmed upon PSF subtraction. All panels are 4\arcsec on the side; North is up and East is left. \label{fig:close_bins}}
\end{figure}


\subsection{PSF Subtraction}
\label{subsec:shane_psfsub}

While ShaneAO provides a high Strehl ratio performance, the light from the bright central star still severely limits our ability to detect faint companions. We therefore proceed to employ a PSF subtraction method to improve the achievable contrast. Because the instrument is not set to enable field rotation (unlike our GPI observations), we decided to use the RDI method, which relies on images of single stars to evaluate the PSF associated with any given observations. Rather than using one PSF at a time and determining which provides the best subtraction, we adopt the Locally Optimized Combination of Images \citep[LOCI,][]{Lafreniere2007}. In short, LOCI uses a large library of single star images and finds the linear combination of these that produces the smallest residuals in the least squares sense. After some testing, we opted to use radial width of 6 and 19 pixels (about 0\farcs2 and 0\farcs6, respectively) for the subtraction and optimization radial lengths.

\begin{figure}
\centering
\includegraphics[width=0.8\columnwidth]{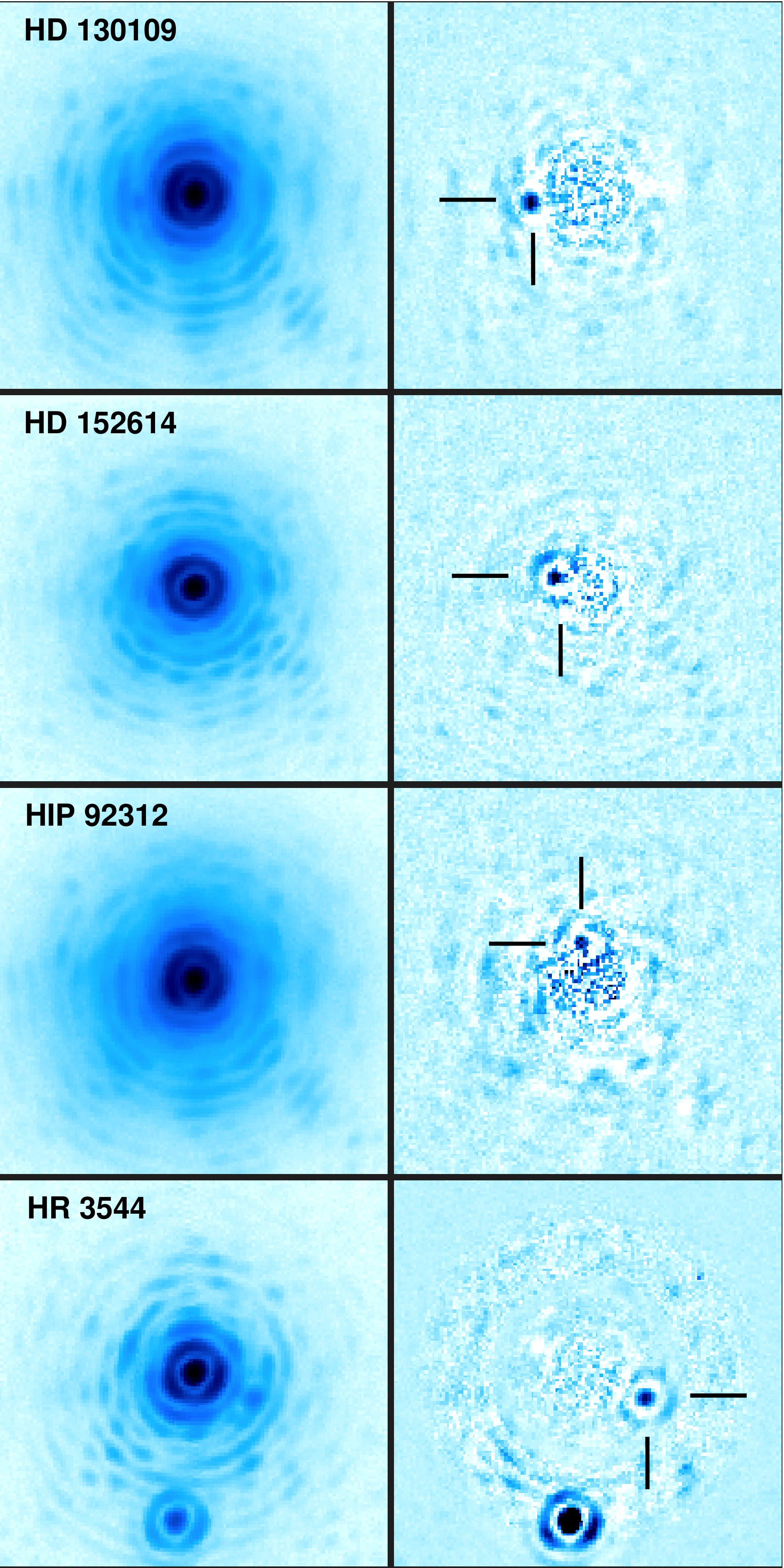}\\
\caption{Companions discovered through the use of LOCI PSF subtraction with ShaneAO observations. The left and right columns represent the direct and LOCI-processed final images of each system, displayed with a logarithmic and linear stretch, respectively. All panels are 4\arcsec on the side; North is up and East is left. \label{fig:loci_bins}}
\end{figure}

Our program is well suited to the use of LOCI as we have roughly 16,500 individual frames to consider in the PSF evaluation. This large volume of data introduces a computing challenge, however, as the LOCI computations employ matrices of size $N \times N$, where $N$ is the number of frames to be considered. To reduce the burden, we pre-computed a ``correlation matrix" of all available frames after applying an 0\farcs7-length median filter to each star-centered frame to remove any residual background light, and normalizing each image to its peak pixel value. In the process of subtracting the PSF for a given frame, we then select the 200 frames with which it is best correlated irrespective of the observing date; frames taken with a different filter or on the same target are excluded from this selection. We then apply the RDI LOCI algorithm and the resulting images are median-combined to produce the final PSF-subtracted image of a target. 

Inspection of the final PSF-subtracted images revealed the presence of an additional five companions that are undetectable, or only marginally detectable, in the corresponding direct images and that are listed in the bottom part of Table\,\ref{tab:shane_bins}. Only the companion to HIP\,116611 is listed in the WDS catalog; the other systems are shown in Fig.\,\ref{fig:loci_bins}. We confirm common proper motion association for the three systems with two epochs of observations (see Fig.\,\ref{fig:common_motion}). We further note that both HD\,130109 and HIP\,92312 have significant Hipparcos-Gaia accelerations consistent with stellar multiplicity \citep{Brandt2021, Kervella2022}, supporting the physical nature of these system. Finally, the companion we discover in the HR\,3544 system makes it a triple system. Except for the companion to HIP\,116611, these LOCI-detected companions are not markedly fainter, but significantly closer (0\farcs3--0\farcs7) from their respective host star, than those found in the direct ShaneAO images. This confirms that the use of LOCI significantly improved the detection limit of our survey, especially at small separations.

In addition to suppressing the starlight, a well-known characteristic of LOCI is to also subtract a fraction of the signal from any companion as the algorithm mistakenly treats it as a local feature in the PSF \citep{Lafreniere2007}, leading to an underestimation of its true brightness. To address this, we proceeded to perform injection and recovery of companions in all images of single stars in the survey. We selected 13 radii ranging from 0\farcs25 to 2\farcs5 and 6 uniformly distributed position angles for these artificial companions. At each radius, we assigned them a flux that corresponds to SNR$\,\approx 15$ based on the detection limits measured in the direct images (see Section\,\ref{subsec:shane_detlim}). We also experimented with fainter artificial companions, but the throughput did not vary by more than 20\%, which is below our estimated uncertainties. Altogether, we generated about 12,000 artificial binaries to evaluate the LOCI throughput for point sources. Each dataset was then processed with LOCI to suppress the PSF and the companion brightness was evaluated by aperture photometry. The throughput was then evaluated by computing the ratio of the apparent brightness to the injected brightness of the companion. 

Given our data acquisition, the vast majority (almost 90\%) of injections were performed in Br$_\gamma$ datasets, which allowed us to evaluate the scatter within individual systems in a given run, as well as the run-to-run scatter of the average throughput curve. We found the two quantities to be comparable at all separation. We thus proceeded to compute a survey-wide throughput curve by computing the median over all datasets in a given filter and, at each separation, assigned the larger of the in-run and run-to-run scatter as the throughput uncertainty. The results are illustrated in Fig.\,\ref{fig:loci_corr}. The curves for both filters are similar, although the $K_s$ images suffer from a marginally higher companion attenuation due to the smoother nature of the images associated with the broadband filter. Within the central $\approx1$\arcsec, the throughput loss is significant, from 0.5 to 4\,mag, confirming that it is mandatory to evaluate it before measuring the brightness of true companions. For each companion detected in PSF-subtracted images, we interpolated the curve of the appropriate filter and applied that correction to the companion brightness as measured by aperture photometry. In addition, we noticed that close companions, inside $\approx$0\farcs5, are pulled up to 0\farcs03--0\farcs04 closer to their primary by the PSF subtraction, indicating that the relative astrometry of the closest LOCI-detected companions are likely inaccurate. Smaller astrometric biases are found further out, with an apparent correlation with the location of the many Airy rings. Given the limited sampling in separation, we do not attempt to correct for this effect and urge caution in using the projected separations listed in Table\,\ref{tab:shane_bins}.

\begin{figure}
\centering
\includegraphics[width=0.8\columnwidth]{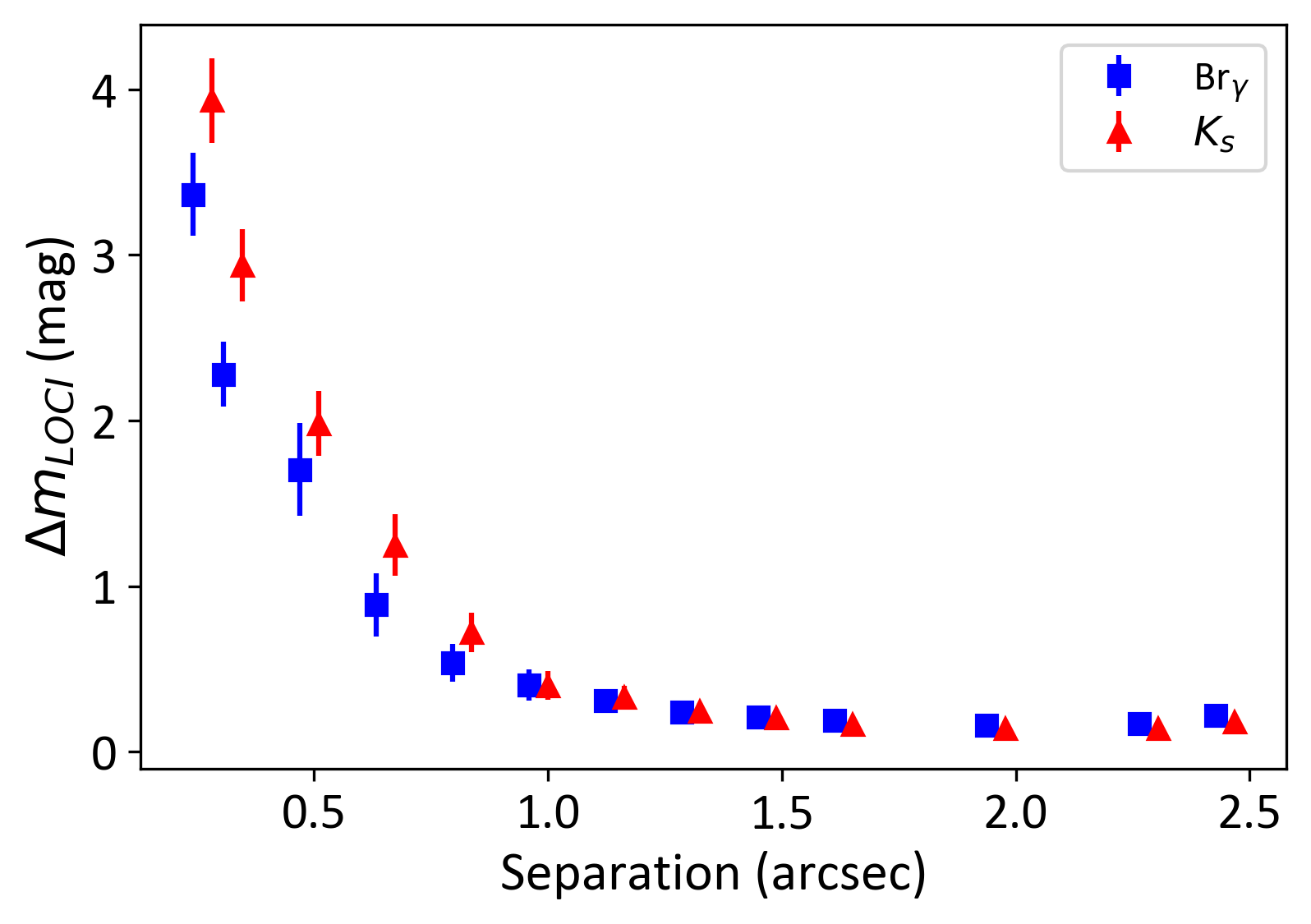}\\
\caption{LOCI throughput evaluated by injection of faint companions in single star datasets and compared the observed and injected companion brightness. The blue squares and red triangles represent the throughput for the Br$_\gamma$ and $K_s$ filters, respectively. To minimize overlap, the curves for the two filters are displaced horizontally by $\pm$0\farcs02 of the true injection separation. Outside of 1\arcsec, uncertainties on the throughput are smaller than the symbol size, or $\lesssim0.1$\,mag. \label{fig:loci_corr}}
\end{figure}


\subsection{Detection Limit}
\label{subsec:shane_detlim}

To determine the detection limit in our survey, we apply the following methodology. Starting with the direct images, we mask detected companions (where relevant), place as many non-overlapping 3-pixel radius aperture as possible in an annulus around the primary, and compute the standard deviation of the resulting flux measurements. We repeat the process for separations ranging from 0\farcs25 to 2\farcs5, and use a 5$\sigma$ detection limit. In the LOCI images, we apply the same process but further apply the throughput correction as determined in Section\,\ref{subsec:shane_psfsub}.

\begin{figure*}
\includegraphics[width=0.8\textwidth]{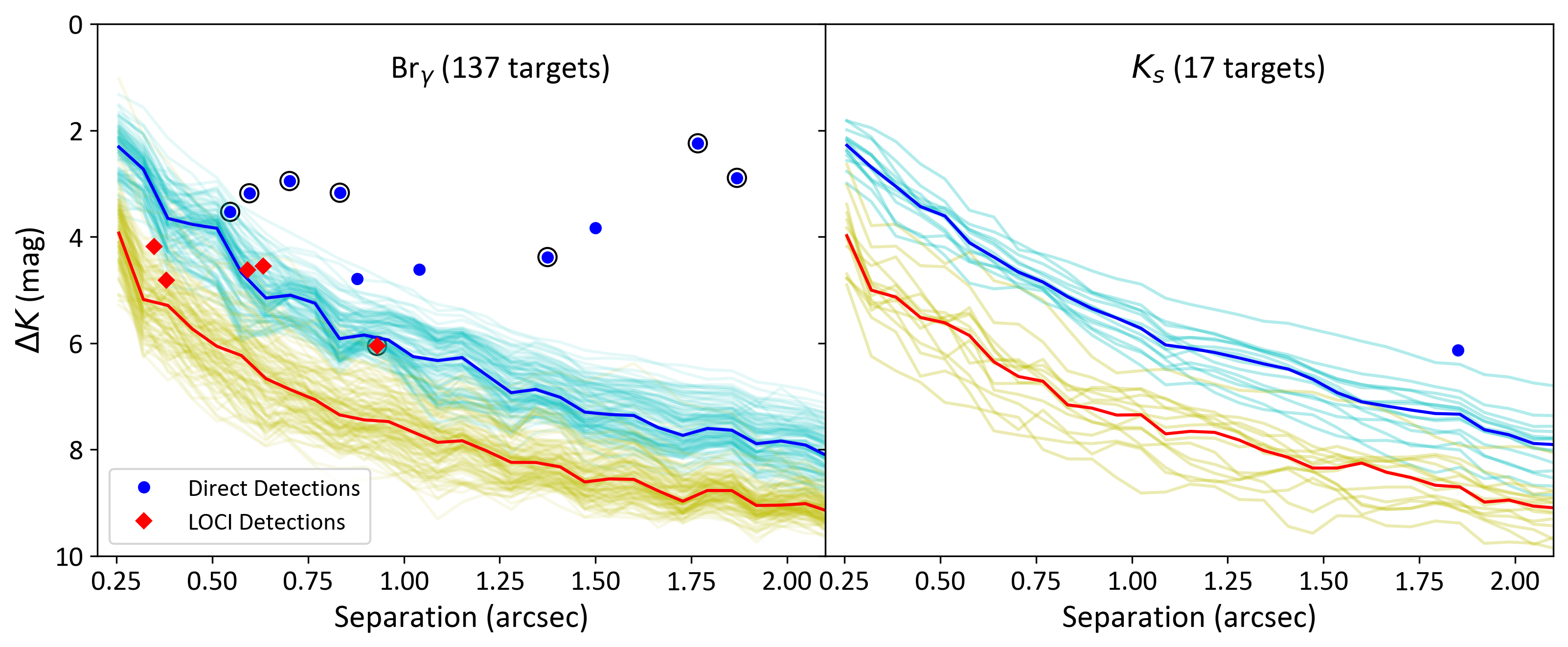}\\
\caption{Detected companions and 5$\sigma$ detection limits in the Br$_\gamma$ and $K_s$ filters for the ShaneAO subsample. Blue circles and red diamonds represent companions that were detected by inspection of the direct images and after LOCI processing, respectively. Circled symbols indicate previously known companions based on the WDS catalog. The thin cyan and gold curves represent the individual detection limits in the direct images and after PSF subtraction, respectively. The solid blue and red curves are the corresponding median detection limits.\label{fig:detect_lims}}
\end{figure*}

The resulting detection curves are shown in Fig.\,\ref{fig:detect_lims} along with all companions detected in either the direct or LOCI-processed final images. The detection limits are marginally better with the Br$_\gamma$ filter, again thanks to the crisper PSF stemming from the narrower bandwidth. Applying LOCI improved the detection limit by about 2\,mag at almost all separations, consistent with the fact that it allowed detection of 5 new companions within the central 1\arcsec. Despite the median LOCI detection limit reaching well beyond $\Delta K = 8$\,mag, the scarcity of companions with $\Delta K \gtrsim 6$\,mag is readily apparent. This suggests a paucity of low-mass stellar companions, which we will analyze quantitatively in Section\,\ref{sec:results}.


\section{GPI Data Analysis}
\label{sec:gpi_analysis}


\subsection{Direct Detections}
\label{subsec:gpi_direct}

Visual inspection of the reduced GPI datacubes revealed 4 clear binaries whose properties are reported in Table\,\ref{tab:gpi_comps}. None of these were known in the WDS catalog prior to our survey. The binary flux ratio is obtained by summing the flux of the companion across all 37 channels and computing the ratio to the same sum for the average of the four satellite spots. The precision of the relative astrometry for these systems is dominated by the location of the (occulted) primary star, which we estimate to be about 0\farcs002. The absolute orientation of the detector is known to 0\fdg2 \citep{DeRosa2020}. Finally, the binary flux ratio precision is conservatively estimated to be 0.05\,mag.

\begin{table*}
	\centering
	\caption{Relative astrometry and photometry of the binary systems detected in GPI images. The sixth column represents the effective temperature derived from the companion brightness and assuming coevality with the primary star, whereas the last two column are the effective temperature and surface gravity that best fit the $H$ band spectra of the companions.}
	\label{tab:gpi_comps}
	\begin{tabular}{cccccccc} 
		\hline
		Target & Obs. Date & Sep. & PA & $\Delta H$ & $T_{\mathrm{eff}}^{\mathrm{phot}}$ & $T_{\mathrm{eff}}^{\mathrm{spec}}$ & $\log g$ \\
		 & (UT) & (\arcsec) & (\degr) & (mag) & (K) & (K) & \\
		\hline
		\multicolumn{8}{c}{Companions identified in direct images}\\
		\hline
HD\,63079 & 2016-01-22 & 0.170$\pm$0.002 & 34.98$\pm$0.70 & 4.61$\pm$0.05 & 3970$\pm60$ & 4400 & 3.5 \\
HD\,63488 & 2016-01-23 & 0.772$\pm$0.002 & 45.96$\pm$0.25 & 4.64$\pm$0.05 & 4220$\pm70$ & 5200 & 3.5 \\
HD\,73287 & 2015-12-25 & 0.184$\pm$0.002 & 157.94$\pm$0.65 & 6.68$\pm$0.05 & 3480$\pm75$ & 3600 & 4.5 \\
HD\,225264 & 2015-12-27 & 0.680$\pm$0.002 & 346.33$\pm$0.25 & 3.06$\pm$0.05 & 4550$\pm150$ & 6600 & 5.5 \\
        \hline
		\multicolumn{8}{c}{Companions identified in PSF-subtracted images}\\
		\hline
HD\,57411 & 2016-01-21 & 0.968$\pm$0.005 & 60.87$\pm$0.36 & 11.75$\pm$0.20 & -- & -- & -- \\
HR\,2986 & 2015-12-26 & 0.607$\pm$0.005 & 131.14$\pm$0.51 & 11.94$\pm$0.20 & -- & -- & -- \\
HR\,3016 & 2015-12-24 & 1.509$\pm$0.005 & 198.81$\pm$0.28 & 11.80$\pm$0.20 & -- & -- & -- \\
        \hline
	\end{tabular}
\end{table*}

In addition to determining the integrated flux ratio of these companions, the GPI datasets provide the $H$ band spectra of the binary flux ratio, which are illustrated in Fig.\,\ref{fig:gpi_comp_spectra}. The companions to HD\,63079 and HD\,73287 have significantly noisier spectra due to their location close to the edge of the coronagraphic mask. Nonetheless, it is clear that all four companions are slightly redder than their corresponding primary, with HD\,73287 being simultaneously the faintest and the reddest of them. This is qualitatively consistent with these objects being physical lower mass companions.

\begin{figure}
\centering
\includegraphics[width=0.8\columnwidth]{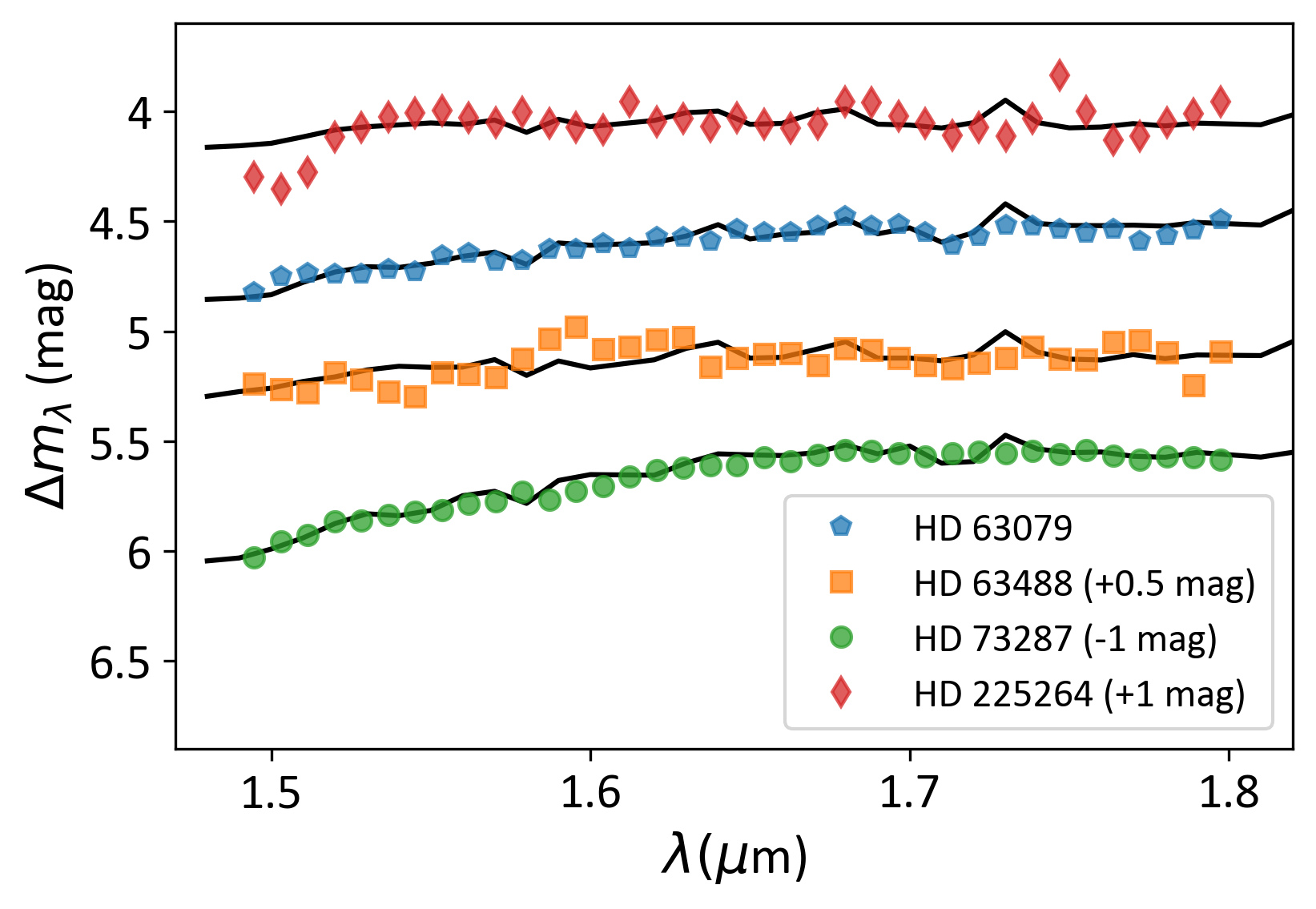}\\
\caption{Flux ratio spectrum of the bright companion detected in the GPI datasets; the spectra have been displaced vertically for visual purposes. The black curves represent the flux ratio spectra obtained by combining the closest Kurucz spectrum and best-fitting NextGen spectrum for the primary and companion, respectively. Departures from a smooth and monotonous flux ratio spectrum are due to individual absorption line, most notably the Br\,10 line at 1.74\,$\mu$m. \label{fig:gpi_comp_spectra}}
\end{figure}

To further analyze the properties of these companions, we perform a fit to the flux ratio spectrum. To this end, we first extract the median effective temperature of the primary star by interpolating the MIST evolutionary models at the median age and mass derived in Section\,\ref{subsec:sample}. Using the flux ratio spectrum measured in the GPI data, we obtain the spectrum of the companion. We then perform a ($T_{\mathrm{eff}}$ -- $\log g$) grid search in the NextGen library of stellar atmospheres \citep{Baraffe1998} where each model spectrum is normalized to the average flux of companion and rebinned to the GPI sampling. In other words, this procedure only aims at fitting the shape of the companion's spectrum, irrespective of its absolute brightness. The best fitting properties are listed in Table\,\ref{tab:gpi_comps} and the corresponding flux ratio spectra are shown in Fig.\,\ref{fig:gpi_comp_spectra}. All fits are satisfactory. We estimate a precision of about 200\,K on the effective temperature of the cooler companions ($\lesssim4500$\,K) but note that once the companions are too hot, their near-infrared spectrum contains too few significant features to be unambiguously characterized. Furthermore, their surface gravity is barely constrained, irrespective of their effective temperature, given the low resolution of the GPI spectra. 

As was the case for some of the ShaneAO-detected companions (Section\,\ref{sec:shane_analysis}), we only obtained one observation of each system. It is therefore impossible to demonstrate that they are physically bound. The reasonable match between photometry- and spectroscopy-based effective temperature for the companions (see Section\,\ref{subsec:comp_mass}) is encouraging but insufficient. Only HD\,63079 has a measurable astrometric acceleration \citep{Kervella2022}, although it is worth noting that the large distance to these targets results in small proper motions and, thus, very small companion-induced accelerations.


\subsection{PSF Subtraction}
\label{subsec:gpi_psfsub}

Although the GPI observing plan was not tailored to enable high-contrast PSF-subtraction, it is still possible to make use of advanced ADI methods for datasets that contain multiple frames even with minimal field rotation, at the price of significant companion self-subtraction. Each dataset was thus processed with the cADI and pyKLIP algorithms \citep{Marois2006, Wang2015}. To enhance detection of any faint candidate companion, the datacubes were collapsed along the wavelength axis, effectively producing broadband images of each system. 

\begin{figure}
\includegraphics[width=\columnwidth]{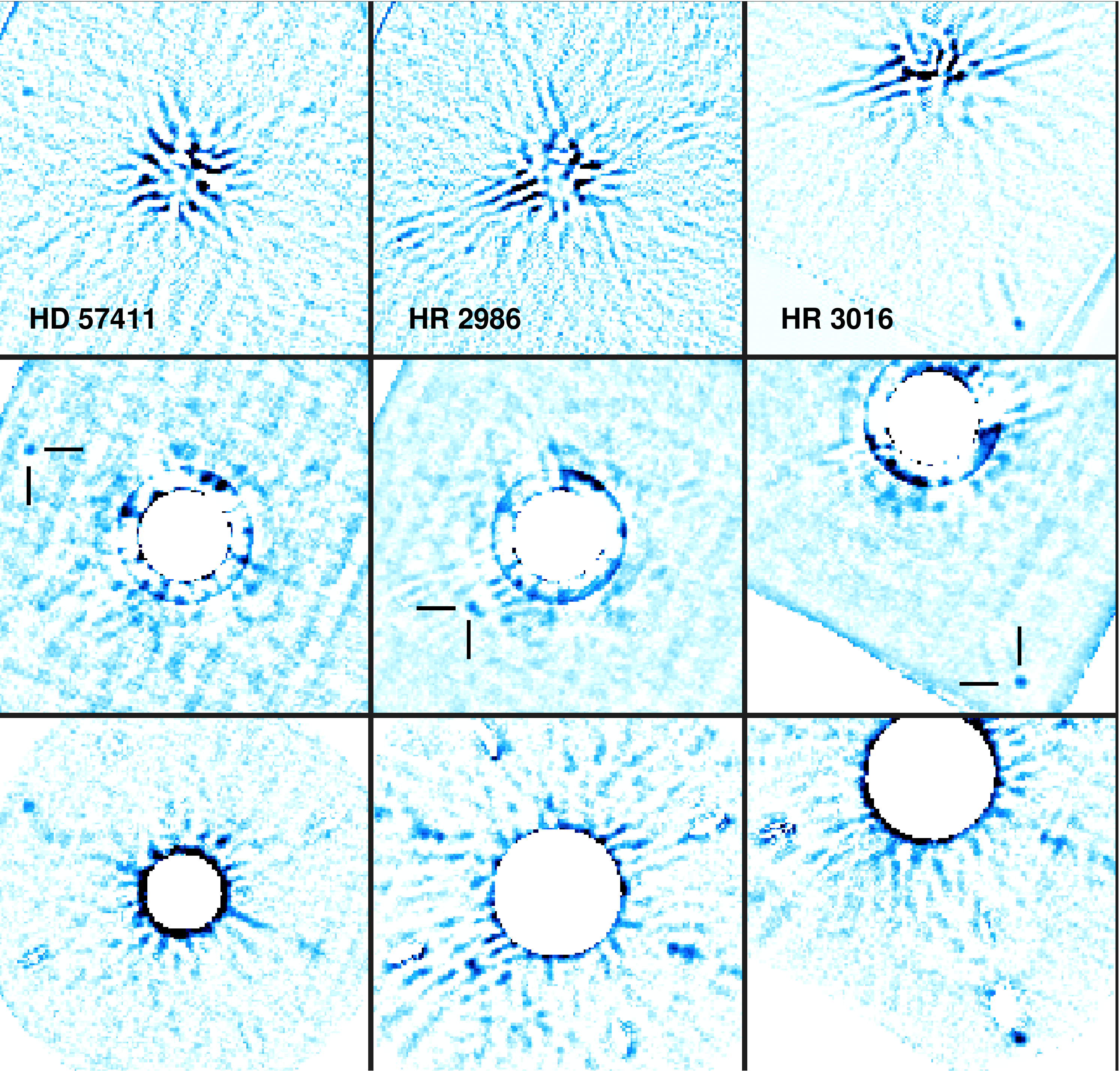}\\
\caption{PSF-subtracted images of the 3 systems in which faint companions were detected. From top to bottom, the rows correspond to the cADI, pyKLIP and mask-and-interpolate PSF subtraction methods. All images are 2\arcsec\ on a side and oriented North up and East to the left. The primary star of HR\,3016 is offset from center to visualize the companion. \label{fig:gpi_faint_comps}}
\end{figure}

Out of 39 datasets that were PSF subtracted, we identified three candidate companions that are detected in both cADI- and pyKLIP- processed images, suggesting that they are true astrophysical sources (Fig.\,\ref{fig:gpi_faint_comps}). We emphasize that, because of the limited field rotation (4--6\degr\ in these systems), we cannot fully rule out that these are residual speckles. Of the three, the candidate companion to HR\,2986 is the least definitive, though it is present with all PSF subtraction methods. To evaluate the location and brightness in as unbiased a way as possible, we apply a third PSF subtraction method, mask-and-interpolate. The method, first applied to GPI data in \citet{Kalas2015}, consists in masking the candidate companion and interpolating with a low-order polynomial the PSF signal from surrounding pixels. This is done on a frame-by-frame basis and the resulting PSF-subtracted images are averaged to produce the final images of the system, shown in the bottom row of Fig.\,\ref{fig:gpi_faint_comps}. While this method is intrinsically less accurate as it relies on interpolation from other pixels, the strict masking of the companions ensure that no companion signal is incorrectly subtracted along the PSF signal. It can therefore provide a less precise, but more accurate estimate of the companion's properties. We thus estimate the location and brightness of all three companions by using a centroid calculation and aperture photometry, respectively. We adopt a conservative uncertainty of 0.2\,mag on the flux ratio. The results are presented in Table\,\ref{tab:gpi_comps}. 

Astrometric acceleration was detected for HR\,2986 and HR\,3016 by \cite{Kervella2022} but required stellar companions to be accounted for, in contrast with the low masses inferred in Section\,\ref{subsec:comp_mass}. No significant acceleration is detected for HD\,57411. In other words, there is no astrometric evidence suggesting that these companions are physically bound.

While PSF subtraction revealed three new candidate companions, it is worth pointing out that we found no companion over a broad range of flux ratio, namely $6.5 \lesssim \Delta H \lesssim 11.5$. This suggests that there truly is a gap in companion brightness, hence mass, for intermediate-mass stars. This is further reinforced by the fact that the faintest ShaneAO-detected companion also lies above this gap. The astrophysical implication of this result requires converting the observed flux ratios into mass ratios, which we present in Section\,\ref{sec:results}, 


\subsection{Detection Limit}
\label{subsec:gpi_detlim}

As with the ShaneAO datasets, we proceed to estimate the detection limits of the GPI observations. Given the high contrast inherent to GPI observations, the contrast required to detect all stellar companions is easily achieved in the reduced images, without further PSF subtraction. Furthermore, both cADI and pyKLIP suffer from very substantial self-subtraction in the case of limited field rotation, which makes it more difficult to correctly evaluate the detection limits in these data products. We therefore estimate the survey's detection limits in the reduced GPI datasets. We employ a similar method as with the ShaneAO datasets. placing non-overlapping apertures in concentric annuli around each target and using 5 times the dispersion of the fluxes measured in them as the detection limit. These limits, as well as the median detection limit across the whole GPI sample, are shown in Fig.\,\ref{fig:gpi_detlim}. As pointed out above, all detected companions are several magnitudes above the survey sensitivity, which is about 10\,mag for separations $\gtrsim$0\farcs5. Conversely, the candidate companions detected through PSF subtraction are well below the initial detection limits, explaining why they were undetected in the reduced images.

\begin{figure}
\centering
\includegraphics[width=0.8\columnwidth]{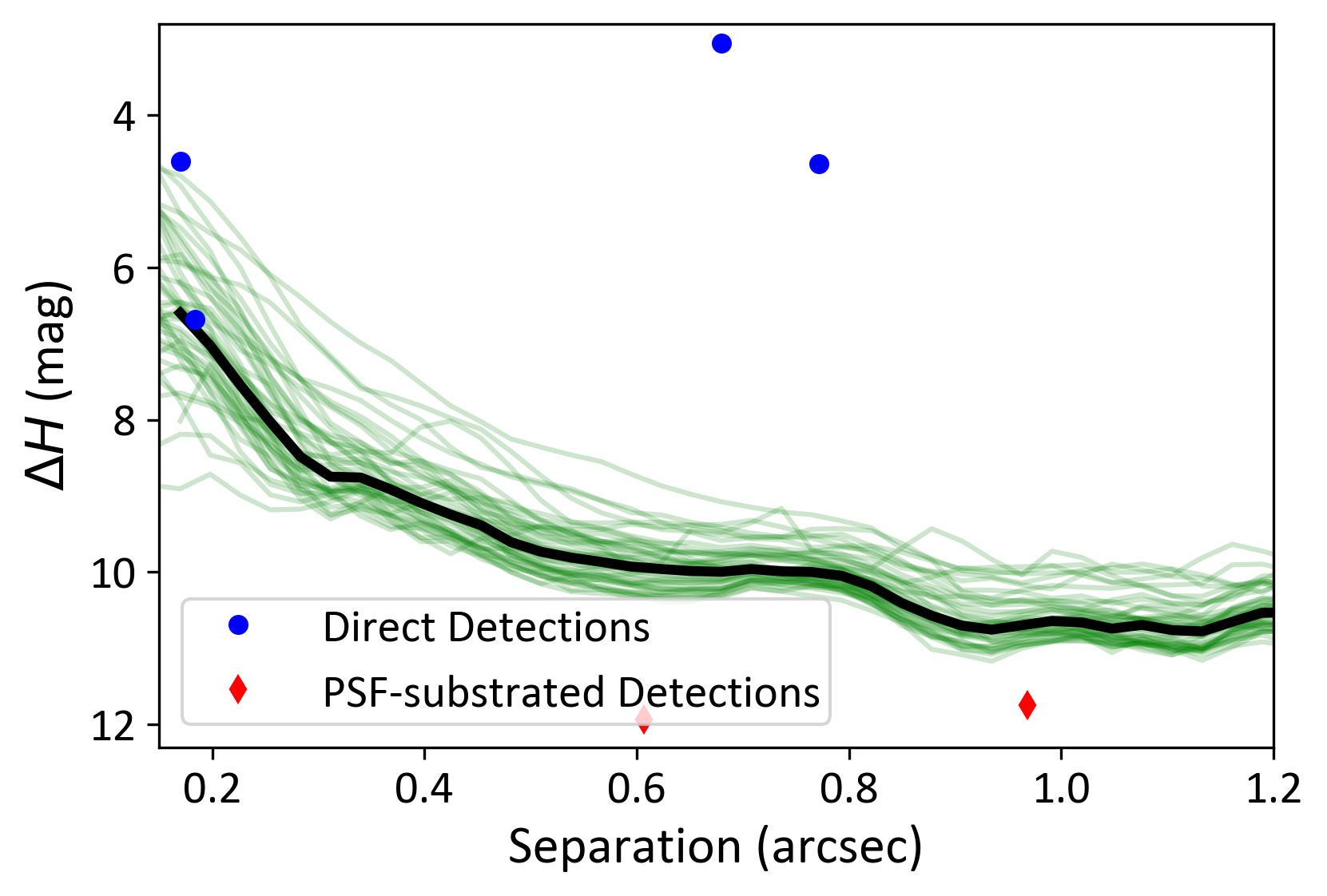}\\
\caption{Detected companions and individual 5$\sigma$ detection limits estimated from the GPI subsample; the black curve represents the median detection limit across the entire survey. Blue circles and red diamonds represent companions detected in the reduced and PSF-subtracted images, respectively. The detection limits are estimated from the reduced images, i.e., without further PSF subtraction processing, as these are sufficient to detect all stellar companions. The faint candidate companion to HR\,3016 is located further out from its primary, where our sensitivity drops due to the square nature of our field-of-view. \label{fig:gpi_detlim}}
\end{figure}


\section{Survey Results}
\label{sec:results}

When combining the ShaneAO and GPI surveys, we identified 24 companions, 16 of which are newly discovered. Among these, 3 are companions to close spectroscopic binaries (HD\,221253, HD\,225264, and HIP\,116611), all with orbital periods $\lesssim6$\,d. Apart from the three very faint companions identified in the PSF-subtracted GPI datasets, all systems have near-infrared flux ratios of $\lesssim6.5$\,mag. We now proceed to evaluate the mass of all companions and to constrain the resulting distributions of companions and mass ratios, with an eye towards possibly identifying a low-mass cutoff to either distribution.


\subsection{Companion Masses}
\label{subsec:comp_mass}

To evaluate each companion's mass, we start from the properties of the primary component based on the MCMC analysis from Section\,\ref{subsec:sample}. We then proceed to generate the isochrone at the appropriate filter ($K$ and $H$ for ShaneAO and GPI, respectively) corresponding to the system's age by interpolation of the MIST evolutionary models. From that isochrone, we interpolate the ``theoretical" magnitude of the primary, use the binary flux ratio (computing the weighted average in case of multiple observations) to obtain the equivalent magnitude of the companion, and interpolate the isochrone to obtain the mass of the companion. We selected this approach over using directly the observed magnitude of the primary as this could be sensitive to unaccounted multiplicity and/or single-filter photometric error. To take into account the various sources of uncertainties in this process, we used a Monte Carlo process whereby we randomly selected a large number (typically 500) of individual MCMC walker position to evaluate the mass and age of the primary star, and tacked on a random Gaussian error on the binary flux ratio. From the ensemble of resulting companion masses, we then report the median and 68-percentile range for all companions in Table\,\ref{tab:masses}. 

The three faint ($\Delta H \gtrsim11.5$) GPI companions are too faint to lie on the MIST isochrones, which have a lowest stellar mass of 0.1\,$M_\odot$. We therefore extended the isochrones to lower masses using the COND models from \citet{Allard2012}. We shifted the COND models by $\approx0.2$--0.3\,mag (depending on age) to ensure they connect smoothly with the MIST models at 0.1\,$M_\odot$. We then applied the same method as described above. All three companions appear to be well into the brown regime (30--50\,$M_\mathrm{Jup}$) based on this analysis. However, we caution that they could also be unrelated background stars. To quantify this possibility, we used the Besan\c{c}on model of stellar populations \citep{Robin2003} to evaluate the density of objects that would fall within 1\,mag of the companion $H$ band brightness in that particular direction in the Galaxy within 1 square degree of their respective target. In all three cases, we find an expected number of background objects of 0.03--0.06 per GPI field-of-view. Given the 43 stars considered in the GPI survey, and assuming we reach similar companion sensitivity (down to $18 \lesssim H \lesssim 19$) in each datasets, we would therefore predict a total of $\sim2$ such unrelated faint background stars. Without proper motion confirmation or spectral information, we cannot conclusively decide the nature of these three objects. Nonetheless, given this statistical analysis, we consider it more likely than not that they are background stars and do not include these candidate companions in our subsequent analysis. For comparison, the same Besan\c{c}on model predicts a density of unrelated field stars with brightness similar to the "bright" GPI companions (Section\,\ref{subsec:gpi_direct}) that is 1--3 orders of magnitude lower, making it highly likely that they are physical systems.

\begin{table}
	\centering
	\caption{Companion masses and system mass ratios (see Section\,\ref{subsec:comp_mass}). The last column indicate whether the system is confirmed as physically bound based on common proper motion (C), likely physically bound based on {\it Gaia}--{\it Hipparcos} acceleration (A), and probably physically bound (P) or most likely background star (b) based on statistical arguments. Systems that include an unresolved spectroscopic companion in the SB9 catalog \citep{Pourbaix2004} are indicated with a $\dagger$ symbol.}
	\label{tab:masses}
	\begin{tabular}{ccccc} 
		\hline
		Target & $M_\mathrm{prim}$ & $M_\mathrm{comp}$ & $q$ & Status \\
		 & ($M_\odot$) & ($M_\odot$) &  & \\
		\hline
		\multicolumn{5}{c}{ShaneAO companions}\\
		\hline
HD\,130109 & 2.32 $^{+0.06}_{-0.05}$ & 0.62 $^{+0.04}_{-0.05}$ & 0.27 $^{+0.01}_{-0.02}$ & A \\
HD\,148112 & 2.59 $^{+0.09}_{-0.10}$ & 1.16 $^{+0.05}_{-0.06}$ & 0.45 $^{+0.02}_{-0.03}$ & C \\
HD\,152614 & 3.11 $^{+0.11}_{-0.11}$ & 0.70 $^{+0.06}_{-0.05}$ & 0.23 $^{+0.01}_{-0.02}$ & C \\
HD\,161734 & 2.29 $^{+0.08}_{-0.08}$ & 0.43 $^{+0.01}_{-0.03}$ & 0.19 $\pm0.01$ & C \\
HD\,204770 & 4.21 $^{+0.17}_{-0.22}$ & 1.92 $^{+0.23}_{-0.21}$ & 0.46 $^{+0.04}_{-0.05}$ & C \\
HD\,221253$^\dagger$ & 4.23 $^{+0.16}_{-0.14}$ & 1.98 $^{+0.22}_{-0.20}$ & 0.47 $\pm0.04$ & C \\
HIP\,3544\,AB & 2.54 $^{+0.10}_{-0.11}$ & 0.74 $\pm0.06$ & 0.29 $^{+0.03}_{-0.02}$ & C \\
HIP\,3544\,AaAb & 2.54 $^{+0.10}_{-0.11}$ & 1.19 $^{+0.05}_{-0.07}$ & 0.46 $^{+0.03}_{-0.02}$ & C \\
HIP\,5310 & 1.76 $^{+0.06}_{-0.09}$ & 0.57 $^{+0.05}_{-0.04}$ & 0.33 $^{+0.02}_{-0.03}$ & A \\
HIP\,13775 & 2.77 $^{+0.08}_{-0.08}$ & 0.67 $^{+0.04}_{-0.02}$ & 0.24 $\pm0.01$ & C \\
HIP\,20648 & 2.13 $^{+0.08}_{-0.08}$ & 1.11 $^{+0.03}_{-0.05}$ & 0.52 $\pm0.02$ & A \\
HIP\,22842 & 2.32 $^{+0.08}_{-0.08}$ & 0.46 $\pm0.02$ & 0.20 $\pm0.01$ & P \\
HIP\,92312 & 2.17 $^{+0.09}_{-0.09}$ & 0.47 $^{+0.04}_{-0.03}$ & 0.22 $\pm0.01$ & A \\
HIP\,107253 & 2.26 $^{+0.09}_{-0.09}$ & 0.87 $\pm0.03$ & 0.38 $^{+0.02}_{-0.01}$ & C \\
HIP\,109745 & 2.21 $^{+0.08}_{-0.08}$ & 0.64 $\pm0.02$ & 0.29 $\pm0.01$ & C \\
HIP\,109831 & 2.21 $^{+0.08}_{-0.09}$ & 0.54 $\pm0.02$ & 0.24 $\pm0.01$ & C \\
HIP\,116611$^\dagger$ & 2.20 $^{+0.09}_{-0.09}$ & 0.28 $\pm0.03$ & 0.13 $\pm0.01$ & C \\
        \hline
		\multicolumn{5}{c}{Bright GPI companions}\\
		\hline
HD\,63079 & 2.56 $^{+0.10}_{-0.08 }$ & 0.56 $\pm0.02$ & 0.22 $\pm0.01$ & A \\
HD\,63488 & 2.24 $^{+0.08}_{-0.08 }$ & 0.63 $\pm0.02$ & 0.28 $\pm0.01$ & P \\
HD\,73287 & 3.33 $^{+0.11}_{-0.11 }$ & 0.33 $\pm0.02$ & 0.10 $\pm0.01$ & P \\
HD\,225264$^\dagger$ & 2.11 $^{+0.08}_{-0.08 }$ & 0.72 $^{+0.02}_{-0.03 }$ & 0.34 $\pm0.01$ & P \\
		\hline
		\multicolumn{5}{c}{Faint GPI companions}\\
		\hline
HD\,57411 & 2.68 $^{+0.09}_{-0.09 }$ & 0.040 $\pm0.002$ & 0.015 $\pm0.001$ & b \\
HR\,2986 & 2.96 $^{+0.10}_{-0.10 }$ & 0.034 $\pm0.003$ & 0.012 $\pm0.001$ & b \\
HR\,3016 & 4.11 $^{+0.13}_{-0.13 }$ & 0.048 $\pm0.004$ & 0.012 $\pm0.001$ & b \\
        \hline
	\end{tabular}
\end{table}

Because we obtained both photometry and spectroscopy of the four bright GPI companions, they provide an opportunity to check the consistency of the resulting stellar properties. If the companions are truly physical, the system's age is well determined, and the evolutionary models are correct, the derived companion masses should correspond to an effective temperature that matches that derived directly from the companion's spectrum. We randomly sampled both the system age from the MCMC chains and the companion masses from the analysis above, and interpolated the MIST evolutionary models to infer the plausible range of effective temperatures for the companions. These are listed in Table\,\ref{tab:gpi_comps} as $T_{\mathrm{eff}}^{\mathrm{phot}}$ and can be compared to the spectroscopic effective temperature, $T_{\mathrm{eff}}^{\mathrm{spec}}$. The two estimates are strongly correlated, although the photometric effective temperature is consistently lower than the spectroscopic one. This suggests the presence of an unaccounted for bias in either (or both) analysis(es). The low resolution of the GPI spectra on the one hand, and imperfections in the derivation of the properties of the primary stars on the other hand, are candidate issues that would require a finer, individual treatment. To ensure consistency across all systems, we adopt the photometry-based properties of the secondary in the remainder of our analysis.


\subsection{Survey Completeness}
\label{subsec:completeness}

In order to analyze the distributions of companion masses and system mass ratios, we must first evaluate the completeness of our survey. To do so, we follow the Monte Carlo methodology outlined in \cite[][and references therein]{Nielsen2019}. For each target, the process can be summarized as follows: we first populate a fine grid of companion mass (or mass ratio) and semi-major axis. At each semi-major axis, we randomly draw a projected separation assuming a random orbit orientation, random orbital phase, and a uniform distribution of orbit eccentricities, appropriate for wide binary systems \citep{Abt2005}. Finally, we assess whether the companion's brightness is above the detection limit for the specific target. For systems with more than one observations we selected the higher contrast datasets for this evaluation. Repeating the process by drawing 1000 projected separations, we then obtain a detection probability map as a function of companion (or mass ratio), semi-major axis and projected separation. Marginalizing over projected separation yields the final completeness map for the target. We repeat the process over all targets and sum of all individual completeness maps produces the final survey completeness maps shown in Fig.\,\ref{fig:tongueplot_mcomp} and \ref{fig:tongueplot_massratio}. One of the advantages of this approach to survey completeness is that it is possible to simply sum the ShaneAO and GPI completeness maps to create a full survey map, as shown in the bottom panel of both figures.

\begin{figure}
\centering
\includegraphics[width=0.8\columnwidth]{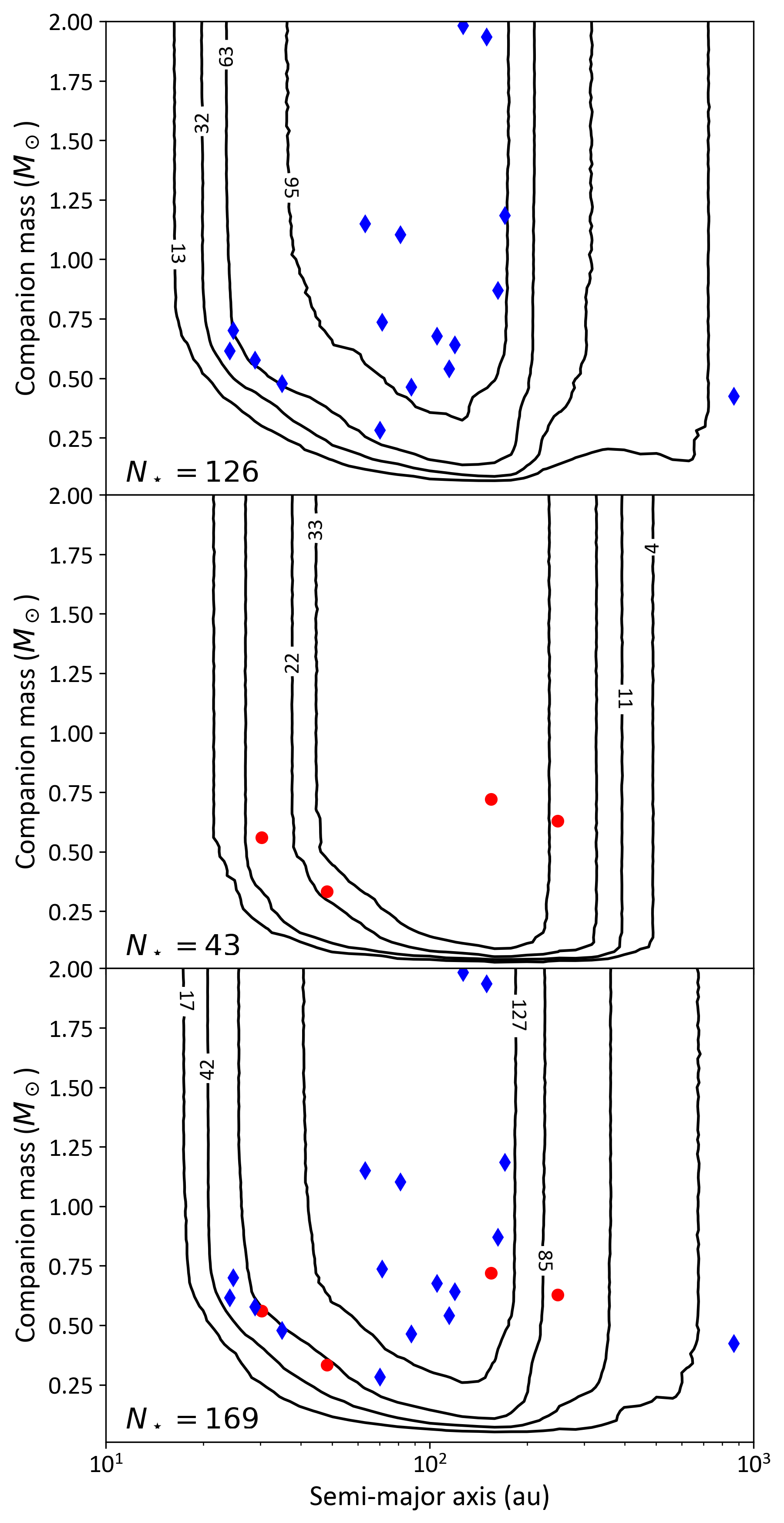}\\
\caption{Survey completeness in terms of companion mass for the ShaneAO, GPI and full survey samples (top, center, and bottom panels, respectively). Each contour marks the sensitivity limit that is obtained for the indicated number of targets; in other words, each contour represents the number of stars for which a companion located above the contour would have been detected in the survey. The contours are set at 10, 25, 50, and 75\% of each subsample size. Companions are plotted at their projected physical separation for lack of information regarding their orbit. The kink at large separations in the lowest ShaneAO contour is induced by the presence of a small number of objects at larger distances. \label{fig:tongueplot_mcomp}}
\end{figure}

Except for a possible higher companion fraction among the lower-mass stars in our sample, we found that splitting our sample by primary mass or by environment (field stars vs cluster members) does not yield statistically significant differences (see Appendix\,\ref{sec:model_subsamples}). For the remainder of this study, we therefore focus on the analysis of the full sample.

\begin{figure}
\centering
\includegraphics[width=0.8\columnwidth]{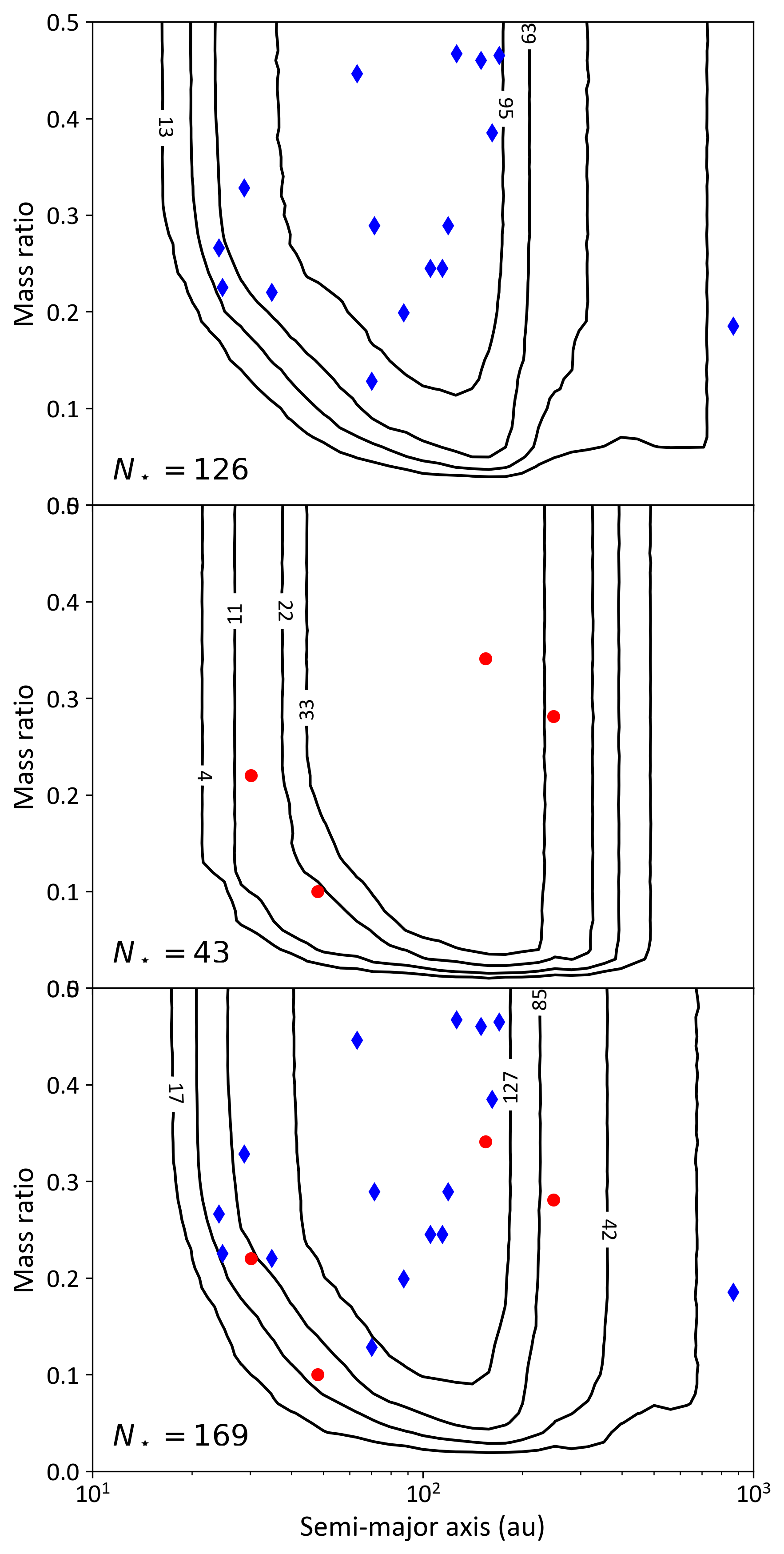}\\
\caption{Survey completeness in terms of mass ratio for the ShaneAO, GPI and full survey samples (top, center, and bottom panels, respectively). Contours and symbols are as in Fig.\,\ref{fig:tongueplot_mcomp}.  \label{fig:tongueplot_massratio}}
\end{figure}

Since the projected separation of a binary is statistical a good approximation for its semi-major axis \citep{Brandeker2006}, the ensemble of their observed positions can be compared to the survey completeness maps. Leaving aside the faint GPI companions, which were only detected through additional PSF subtraction, almost all companions lie above the contours corresponding to 30\% of the (sub)sample size, as we would expect. The companion to HD\,161734, which lies at a projected separation of 870\,au was detected because the source is among the most distant considered it, at 460\,pc. In general, our survey is sensitive down to $\approx0.1\,M_\odot$, or mass ratios of $q\lesssim0.05$ with both ShaneAO and GPI, which satisfies our original goal of distinguishing between a lower limit in companion mass (close to the substellar regime) or in mass ratio (at $q\approx0.1$). Despite this sensitivity, it appears that there is a deficit of systems with $M \lesssim 0.25$ and $q \lesssim 0.1$, which we explore quantitatively in the next subsection.


\subsection{Modeling the Distributions of Companion Masses and Mass Ratios}
\label{subsec:model_distrib}

Combining the companions we have detected with the survey completeness maps, we can now proceed to model the distribution of companions to intermediate-mass stars, following the formalism used by \cite{Nielsen2019}. Since we are interested in assessing whether that population is characterized by a lower cutoff in companion mass or in mass ratio, we perform a similar modeling for both quantities. Specifically, we assume that companions follow a power law distribution of semi-major axes with an index $\beta$, where $\beta = 0$ corresponds to a log-uniform distribution, commonly known as \"Opik's law. Similarly, we assume power law distributions for the companion masses or the mass ratio, with indices $\alpha_\mathrm{m}$ and $\alpha_\mathrm{q}$, respectively. We also consider the minimum companion mass or mass ratio, $M_\mathrm{min}$ and $q_\mathrm{min}$, as an additional free parameter. Finally, the distributions are normalized through the companion fraction over the parameter space under consideration, $f$. Overall, our model has four parameters, which we explore through a densely populated grid over broad ranges of values, as described in Table\,\ref{tab:models}, with upper limits on $M_\mathrm{min}$ and $q_\mathrm{min}$ set by the lowest mass companion (or lowest mass ratio system) identified in the survey.

\begin{table}
	\centering
	\caption{Binary population model parameters and best fit values. Lower limits indicated in the third column represent 3$\sigma$ confidence levels.}
	\label{tab:models}
	\begin{tabular}{ccc} 
		\hline
		Parameter & Explored Range & Best Fit \\
		\hline
		\multicolumn{3}{c}{Fit to companion mass distribution}\\
		\hline
$\beta$ & [-3 .. 3] & -0.16 $\pm0.28$ \\
$\alpha_\mathrm{m}$ & [-3 .. 3] & -1.32 $\pm0.39$ \\
$M_\mathrm{min}\ (M_\odot)$ & [0.08 .. 0.25] & $\geq 0.21$ \\
        \hline
		\multicolumn{3}{c}{Fit to mass ratio distribution}\\
		\hline
$\beta$ & [-3 .. 3] & -0.05 $\pm0.22$ \\
$\alpha_\mathrm{q}$ & [-3 .. 3] & -0.74 $\pm0.44$ \\
$q_\mathrm{min}$ & [0.01 .. 0.10] & $\geq 0.073$ \\
        \hline
	\end{tabular}
\end{table}

To constrain the free parameters given the companions found in our survey, we first rebin all known companions into a grid that samples mass (or mass-ratio) linearly and projected separation log-linearly. For each combination of model parameters, the 3-dimensional completeness map of the survey (as a function of companion mass or mass ratio, semi-major axis, projected separation) is weighted based on the model distributions, which describes how companions are distribution in terms of physical parameters. We then marginalize over the semi-major axis dimension to produce a binned distribution of expected companions that can be directly compared to the observed distribution. In each bin, we use the Poisson likelihood and compute the product over all bins to obtain a final probability that the set of model parameters can produce the observed distribution. In a Bayesian formalism, we then multiply by priors on each model parameters to infer their posterior distributions. We assume uniform priors on $\beta$, $\alpha_\mathrm{m}$, $\alpha_\mathrm{q}$, $M_\mathrm{min}$, and $q_\mathrm{min}$, and Jeffrey's prior on the companion frequency, Prior($f$)$ = 1 /\sqrt{f}$, to match the assumption of a Poisson process. The latter parameters is the least important one for the present study and has essentially no influence on the results for the other parameters.

\begin{figure}
\centering
\includegraphics[width=0.8\columnwidth]{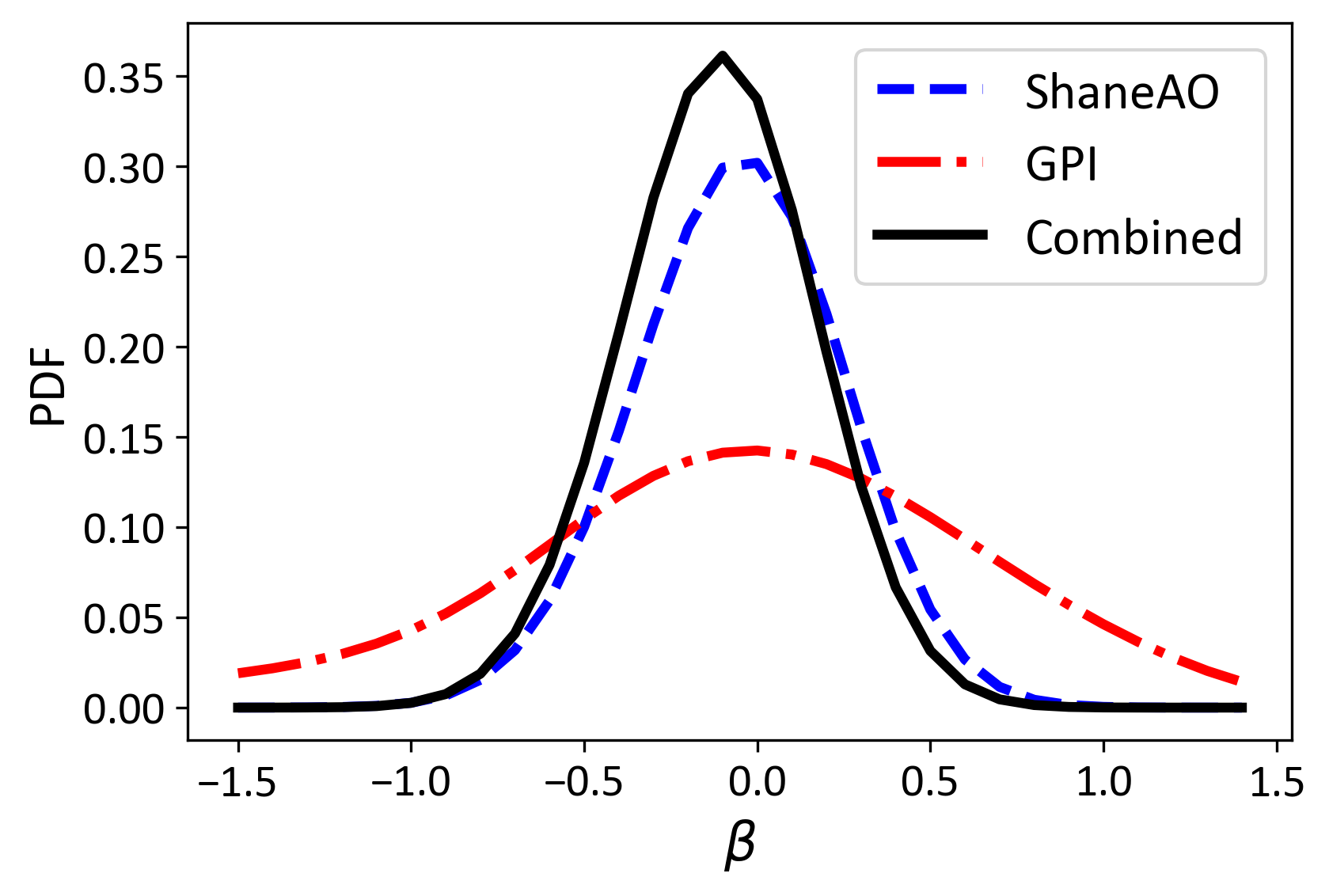}\\
\caption{Posterior distribution on the semi-major axis power law index, $\beta$, based on fitting the distribution of companion masses; The results from the fit to the companion mass ratio distribution are undistinguishable. The dashed blue, dot-dashed red and solid black curves represent the fit to the ShaneAO, GPI and combined samples, respectively. \label{fig:bayes_beta}}
\end{figure}

\begin{figure*}
\includegraphics[width=0.45\textwidth]{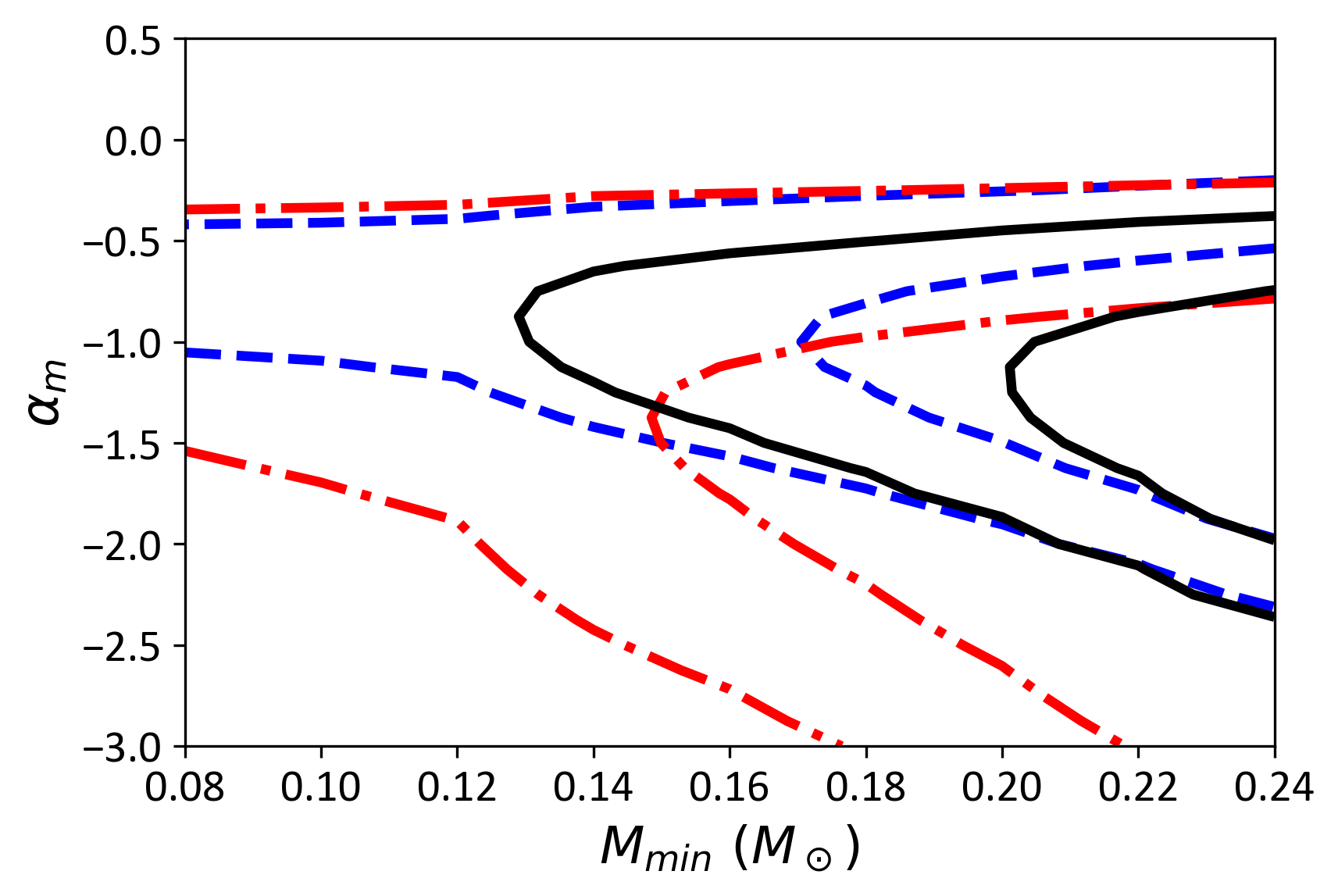}\hspace*{0.5cm}
\includegraphics[width=0.45\textwidth]{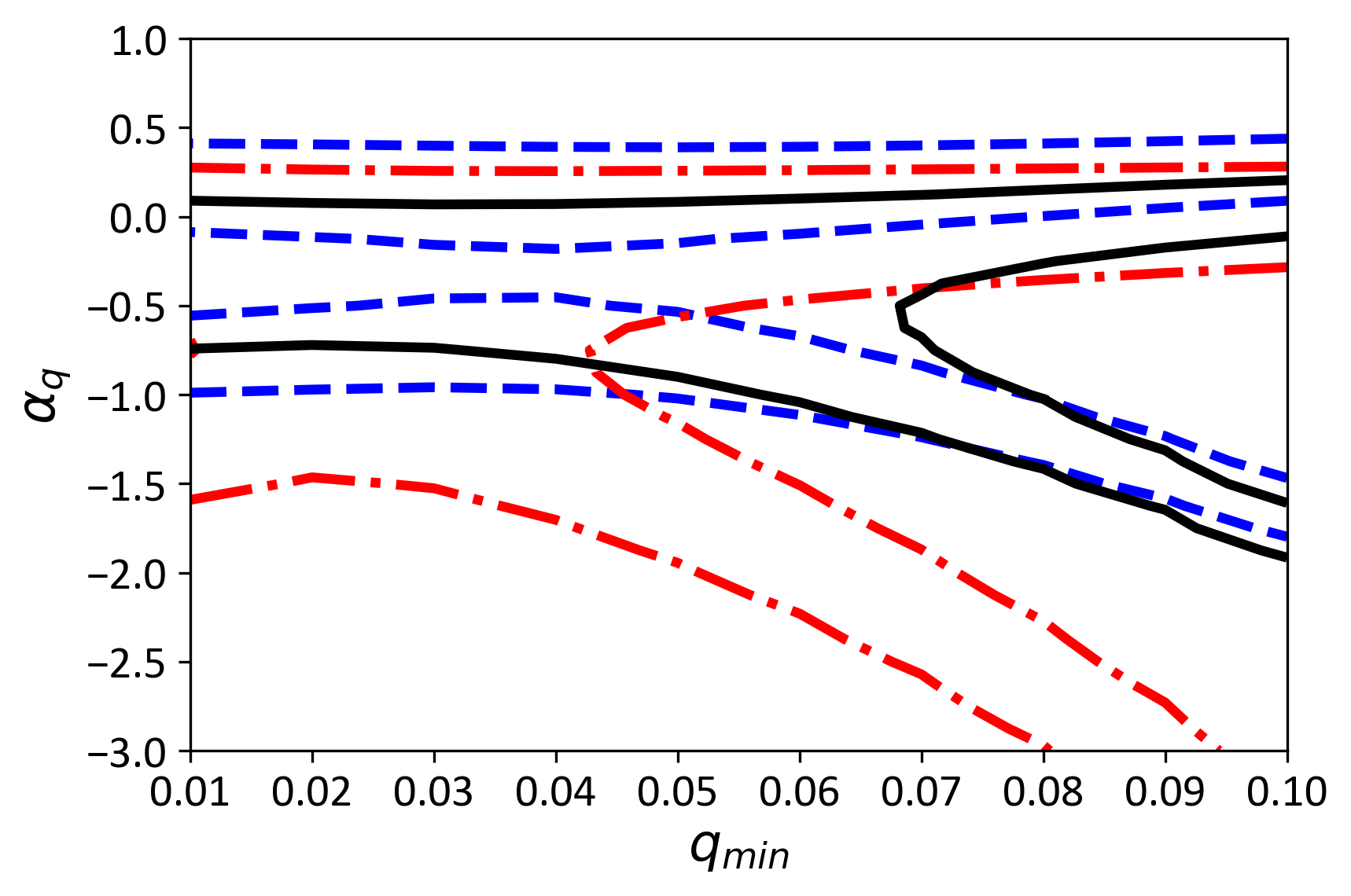}\\
\caption{{\it Left:} Posterior distributions on the companion mass power law index and minimum companion mass for all companions below 1\,$M_\odot$.  The dashed blue, dot-dashed red and solid black contours represent the 68 and 95\% confidence level from the fit to the ShaneAO, GPI and combined samples, respectively.  {\it Right:} Posterior distributions on the companion mass ratio power law index and minimum companion mass ratio for all systems with $q\leq0.5$. \label{fig:bayes_alpha}}
\end{figure*}

Because our survey excluded bright companions that would have negatively affected our ability to detect companions, we cannot constrain the entire distribution of companion masses or mass ratios. Furthermore, there is significant evidence that a single power law for either quantities is likely insufficient to reproduce the overall observed distributions \citep[e.g.,][]{Moe2017}. Instead, we consider only the low-mass end of the distributions. Specifically, we only consider companions with masses $M \leq 1\,M_\odot$, or $q \leq 0.3$, respectively. This upper mass ratio limit is consistent with the treatment of \citet{Moe2017}. We also tested using limits of $M \leq 2\,M_\odot$ or $q \leq 0.5$ and found indistinguishable results.

Posterior distributions for the model parameters are presented in Fig.\,\ref{fig:bayes_beta} and \ref{fig:bayes_alpha}. Due to a smaller sample size, the constraints from the GPI subsample are weaker than the ShaneAO one, but the two are consistent with one another for all parameters of interest. We therefore only discuss the results from the fit to the combined sample. We find that the semi-major axis distribution is close to \"Opik's distribution, a result of the spread of companions across the separation range we sampled. The lone ShaneAO companion at $\approx900$\,au projected separation, outside even the 10-star completeness curve, could be biasing slightly the distribution, although the effect is expected to be small. Furthermore, the presence of  a significant fraction of companions close to the 70-star contour plots at small semi-major axes (see Fig.\,\ref{fig:tongueplot_mcomp} and \ref{fig:tongueplot_massratio}) likely compensates for this outlier. The posterior distributions for $\alpha_\mathrm{m}$ and $\alpha_\mathrm{q}$ are well constrained, with significantly negative values for both parameters, again likely because of the presence of many companions in the lower-left region of the completeness maps. 

The parameters of highest interest for this study are $M_\mathrm{min}$ and $q_\mathrm{min}$. In both cases, the model fit favors the highest values that we probed. Values beyond the explored range are excluded since at least one companion would then violate the model. Unsurprisingly, there is a correlation between $M_\mathrm{min}$ ($q_\mathrm{min}$) and $\alpha_\mathrm{m}$ ($\alpha_\mathrm{q}$), with lower values of the companion cutoff associated with shallower distributions. Given the modest number of companions and the limited parameter space between the lowest mass companions and the survey sensitivity limit, it is not possible to exclude any combination of parameters at the 3$\sigma$ level. Nonetheless, we consider the results of the model fit as tantalizing evidence for a dearth of companions with $M \lesssim 0.25\,M_\odot$, or systems with $q \lesssim 0.1$, i.e., a deficit of ``extreme" binaries among intermediate-mass stars. We discuss the implications of these results in Section\,\ref{sec:discus}.


\section{Discussion}
\label{sec:discus}


\subsection{Comparison With Past Surveys}

As described in Section\,\ref{subsec:completeness}, our survey achieved its goal of being sensitive to stellar companions all the way down to the substellar limit over about a decade in separation ($\approx30$--300\,au) for the majority of our targets. While the sample construction purposely excluded higher mass companions, which were well characterized for most targets, we identified two dozen low-mass stellar companions. This allowed us to constrain their ensemble statistical properties (Section\,\ref{subsec:model_distrib}) which we know compare to previous surveys of intermediate-mass stars binaries.

We find that the distribution of semi-major axis is consistent with \"Opik's (log-uniform) law, consistent with results from surveys in the young Scorpius-Centaurus association \citep{sha02,kou07}. Among field stars, \cite{der14b} found a broad log-normal distribution peaking at $\approx400$\,au, with a modest rise from 30 and 1000\,au that is likely consistent with a log-uniform distribution over that range. Conversely, \cite{ElBadry2019} found a significantly declining distribution, with $\beta\approx-1$ (this index is listed as $\gamma_a$ in their study), but their survey has limited sensitivity to separations $\lesssim300$\,au. This may be further indication of the turnaround in the separation distribution that \cite{der14b} interpreted as the tail of a log-normal function. In our survey, there is a tantalizingly smaller number of companions towards the outer 25--50\% sensitivity range than towards the corresponding inner range, suggesting that the distribution of separation drops outside of $\approx300$\,au, but the number of companions involved is too low to be conclusive. Ultimately, the separation distribution must be characterized over a much broader range than covered in our survey to explore departures from a pure power law distribution, but we conclude that our survey results are broadly consistent with past surveys.

Both the distributions of companion masses and of mass ratios are characterized by a significantly negative power law index ($\alpha \approx-1$) and a minimum cutoff corresponding to a companion mass of 0.2--0.25\,$M_\odot$. These power law indices are reasonably consistent with those found in the literature, with $\alpha_q \approx -0.5$ estimated when fitting the entire distribution of mass ratios down to $q=0.1$ \citep[][excluding companions outside of 125\,au in the latter survey]{reg11,der14b}. We note that \cite{ElBadry2019} suggested that $\alpha_q \gtrsim 0$ and $\alpha_q \approx -1$ for companions above and below $q = 0.3$, respectively, forming a clear peak at this break point in mass ratios. Like for the separation distribution above, our analysis is hardly proof that the distribution indeed follows a power law given the limited range of mass ratios considered here. Nonetheless, our survey provides strong support for a turnaround in these distributions, from slowly rising towards lower mass companion to a sharp decline that makes extreme mass ratio binaries rare among intermediate-mass stars.


\subsection{A Low Mass Ratio Desert}

\begin{figure*}
\includegraphics[width=0.8\textwidth]{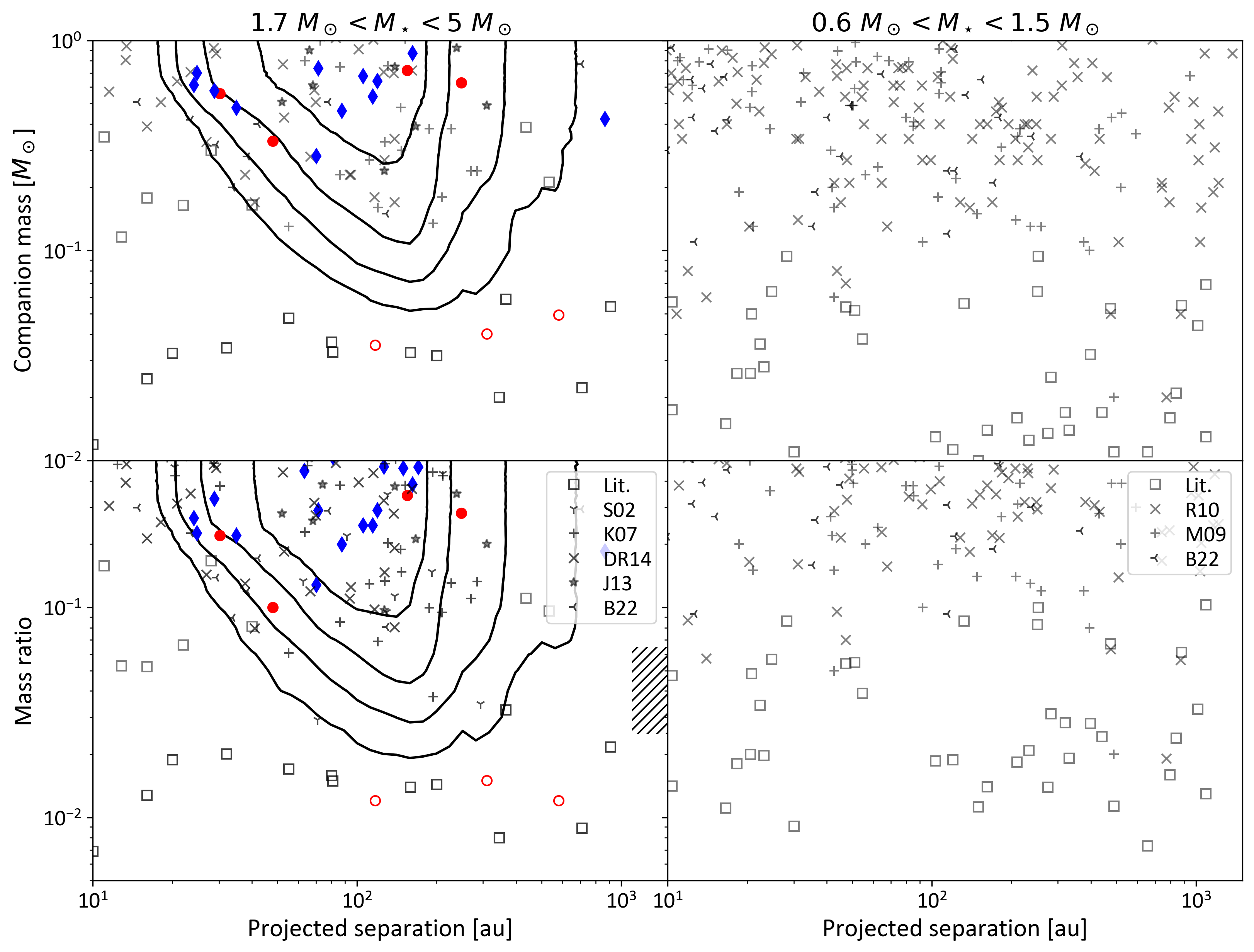}\\
\caption{{\it Left:} Survey completeness contours and binary systems detected in this survey. Blue diamonds and red squares represent stellar companions detected in our ShaneAO and GPI surveys, respectively. Notice the logarithmic scale along the vertical axis to better visualize the ``extreme" binaries regime. The empty red circles indicate the faint candidate companions detected in PSF-subtracted GPI datasets that are most likely unrelated background stars. Gray symbols represent companions to intermediate-mass stars from the \citet[][S02]{sha02}, \citet[][K07]{kou07}, \citet[][J13]{Janson13}, \citet[][DR14]{der14b} and \citet[][B22]{Bonavita2022b} surveys. Only companions with separations of 2\farcs5 or less are considered to reduce contamination by unrelated background sources. Squares indicate individual ``extreme mass ratio" systems in the literature (see Appendix\,\ref{sec:extreme_lit}). The hatched rectangle indicates the approximate range of mass ratios of the ``companion desert." {\it Right:} Distribution of companions to solar-type stars from the \citet[][R10]{Raghavan2010}, \citet[][M09]{met09}, and \citet[][B22]{Bonavita2022b} surveys and substellar companions from the literature (Appendix\,\ref{sec:extreme_lit}). \label{fig:q_extreme}}
\end{figure*}

The sharp lower limit (in companion mass or mass ratio) that we identified in this survey must be interpreted by considering the wider context. We first focus on intermediate-mass stars, among which a number of ``extreme" binary systems has been identified in the past. In most cases, the companion is firmly in the brown dwarf regime, but a few low-mass stellar companions are sometimes placed in that category as well. The left column of Fig.\,\ref{fig:q_extreme} places the results of our survey in this broader context.

Focusing first on systematic surveys \citep{sha02, kou07, Janson13, der14b}, we immediately notice the almost complete dearth of companions with $q\lesssim 0.08$ if we exclude systems wider than 2\farcs5 in projected separation, similar to our outer search radius. The motivation to exclude wider companions is not merely to match methodology, but to reduce the risk for confusion with unrelated background stars. The population of wider companions identified in these surveys typically contain many candidate companions with mass ratios as low as $q\approx0.05$ but whose physical nature is uncertain. To minimize this issue, \cite{kou05} placed a brightness limit on companions that effectively excludes companions with $M \lesssim 0.3\,M_\odot$ and \cite{der14b} only considered companions with $q \geq 0.15$ in their statistical analysis. In other words, while past surveys had found candidate ``extreme" companions, their reliability was doubtful. Our survey does not identify any new companion below $\approx0.3\,M_\odot$, only modestly higher than the lowest mass companions found in other surveys ($\approx0.15\,M_\odot$). This is strong evidence for a lower companion mass cutoff to the stellar binary population.

The rapidly growing population of physically-confirmed ``extreme" binaries found through high-contrast imaging (listed in Appendix\,\ref{sec:extreme_lit}) is also shown in Fig.\,\ref{fig:q_extreme}. Because most of these were not discovered through uniform surveys, it is impossible to derive a robust companion mass (or mass ratio) distribution. Nonetheless, this population appears split between two regimes: low-mass stellar companion down to $\approx0.15\,M_\odot$ and brown dwarf companions in the 10--50\,$M_\mathrm{Jup}$. In other words, there is a clear demarcation between a binary-like population that extends smoothly from a lower-mass limit all the way to equal mass systems on one hand, and a substellar population that has a maximum companion mass (or mass ratio) on the other hand. From Fig.\,\ref{fig:q_extreme}, we find that the $\approx0.05$--0.15\,$M_\odot$ and $q \approx 0.02$--0.05 ranges are the most devoid of companions. We propose that these ranges be referred to as a ``low-mass companion desert" to intermediate-mass stars.

To interpret the physical nature of this new desert, we must compare it to its solar-type counterpart, the brown dwarf desert. To this end, we assembled all companions identified in the \cite{met09} and \cite{Raghavan2010} surveys, supplemented by all ``extreme" companions we found in the literature (see Appendix\,\ref{sec:extreme_lit}). These are shown in the right column of Fig.\,\ref{fig:q_extreme}. As discussed above, the brown dwarf desert is not completely dry among visual companions and there are a number of substellar companions now known to orbit solar-type primaries. Nonetheless, as discussed in \cite{reg11}, the $\approx 0.02$--0.05\,$M_\mathrm{Jup}$ range is the least populated. Conversely, this is precisely the range in which most substellar companions to intermediate-mass stars have masses are found.

As Fig.\,\ref{fig:q_extreme} illustrates, the least populated ranges for intermediate-mass and solar-type stars line up much better when considering mass ratio as the quantity of interest. In other words, while the populations of companions to both intermediate-mass and solar-type stars are characterized by a distinct desert, it can only be represented as a unique feature independent of stellar mass if it is an attribute of the mass ratio distribution. We therefore conclude that the true nature of the brown dwarf desert, as well as of the low-mass companion desert identified here for intermediate-mass stars, is a consequence of a ``fixed (lower) mass ratio limit" for stellar binaries. Given the analysis presented here, we evaluate that this lower limit is $q_\mathrm{min}\approx 0.05$--0.075.


\subsection{Implications}

The finding that intermediate-mass stars possess their own low-mass companion desert and that it corresponds to the same range of mass ratios as the brown dwarf desert that is associated with solar-type stars has a number of observational and theoretical implications. 

First of all, it further strengthens the notion that the mass ratio is a more fundamental property of multiple systems than the masses of individual components \citep{Goodwin2013}. In this context, we caution against studies of the demographics of substellar companions that combine stars of different mass and parametrize the distribution as a function of companion mass \citep[e.g.,][]{bra14, Nielsen2019} as this could blur important distinctions such as the universal mass ratio desert identified here. Instead, we assert that brown dwarf companions to intermediate-mass stars are the counterparts to directly imaged giant planets to solar-type stars rather than to the few objects in the brown dwarf desert. This realization likely explains the fact that the vast majority of planetary-mass companions that have been detected to date are associated with primary stars with $M_\star \gtrsim 1.5\,M_\odot$ \citep{Bowler2016, Nielsen2019, Vigan2021}. Assuming that mass ratio is the most relevant quantity, the counterpart of the 5--10\,$M_\mathrm{Jup}$ planets around intermediate-mass stars would instead be in the 1--5\,$M_\mathrm{Jup}$ range for solar-type stars, which is only marginally probed by even the most recent dedicated surveys \citep{Nielsen2019, Vigan2021}. We therefore predict that future higher-contrast instruments will identify a rapidly increasing number of planets around solar-type stars as they probe fainter companions. For the same reason, a uniform deficit of $q\approx0.02$--0.05 systems preferentially deprives solar-type stars from substellar companion, thus explaining the observed positive correlation between the occurrence rate of such companions and stellar mass as a Malmquist-like bias \citep[e.g.,][]{Baron2019, Nielsen2019}.

Conversely, if the same companion desert also applies to lower-mass stars, which are believed to form in a similar manner, we expect that they should host a substantial fraction of brown dwarf companions, corresponding to $q\approx0.1$--0.5 mass ratios. A few such systems have already been identified \cite[e.g.,][]{nak95, Rebolo1998, Luhman2006} although their statistical analysis is affected by the generally old (and often highly uncertain) age of M dwarfs, which leads to substellar companions being extremely faint. The occurrence rate of such companions is still an open question being actively investigated \citep{Bowler2015, Lannier2016, Gauza2021, Salama2021}. At shorter separation, giant planets (and by extension brown dwarfs) are much rarer than among solar-type stars \citep[e.g.,][]{Fulton2021}, in line with the trend we predict at wider separations. At the other extreme, the presence of a gap in the distribution of (stellar) companions remains to be determined \citep[e.g.,][]{Reggiani2022} and is complicated by the large distance and extreme contrast between primary and companion stars. Besides, it is likely that the formation of massive stars involves more complex processes and therefore, the presence of a universal mass ratio gap for low-mass up to intermediate-mass stars does not guarantee that it should apply at for the highest mass stars.

From a theoretical perspective, the low-mass companion desert further reinforces the notion that there are two mechanisms that lead to the formation of extreme binaries. The lowest mass companions (within a few hundred au) are believed to form within circumstellar disks whereas the higher-mass ones most likely form through cloud fragmentation like stellar binaries. Given that the relationship between the masses of young stars and their circumstellar disks is roughly linear \citep{Andrews2013, Barenfeld2016, vanderMarel2021}, it is natural to expect that the companion mass ratio distribution, including its high-mass tail, that results from disk-mediated processes would be essentially independent of stellar mass. For instance, standard core accretion models of planet formation predict a linear dependency on the occurrence rate of giant planets with stellar mass \citep[e.g.,][]{Laughlin2004,Alibert2011}. Similarly, \cite{Kratter2010} predicted that intermediate-mass stars would host more brown dwarf companions than solar-type stars if gravitational instability in disks is a significant formation channel. While quantitatively testing the predictions of such models is beyond the scope of this study, the results presented here are in qualitative agreement. 

We now turn our attention towards the low-mass end of the stellar binary regime which we expect to be associated with cloud fragmentation. Our analysis shows that this mechanism is characterized by a fixed lower mass ratio limit, at $q_\mathrm{min}\approx 0.05$--0.75. Therefore, not only is the overall mass ratio distribution largely independent of the primary mass \citep{reg11}, but so is its lowest limit. This suggest that cloud fragmentation is a mostly scale-invariant process, with higher mass prestellar cores forming higher mass stars that have a similar mass-ratio distribution to lower-mass cores. This is intuitively consistent with a scenario in which most of the final mass of the systems' components are accreted after the initial fragmentation, i.e., a situation where the self gravity and structure of cloud is the main driver of the final mass ratio. In this context, we note that numerical models of cloud fragmentation generally find modest or no dependency of the mass ratio distribution with stellar mass \citep[1.g.,][]{Goodwin2004, Bate2009}, consistent with this scenario. Similarly, the observed distribution of prestellar core masses points to a linear relationship between their mass and that of the stars they form, suggesting a largely scale-invariant process for cloud collapse and fragmentation \citep{Chabrier2010, Guszejnov2015}, although the treatment of multiplicity, which is a strong function of stellar mass, adds complexity to this simple analysis \citep{Holman2013}.

Lastly, we note that the invariance of mass ratio distribution with stellar mass appears to break down at large separations \citep[$\gtrsim1000$\,au][]{ElBadry2019}, with a possibly larger occurrence rate of extreme binaries among intermediate-mass stars \citep[e.g.,][]{sha02, der14b}. This is an indication that the formation mechanism and/or dynamical evolution of these wider systems is different from the visual systems under consideration in this study. It could be that if cloud fragmentation leads to an initially large separation, most of the subsequent accretion occurs on the more massive primary star. Alternatively, the system may have form as a tighter pair but the companion migrated outwards, to a low-density region of the cloud where little mass is available. One such scenario calls for the ``unfolding" of initially non-hierarchical triple systems \citep{Reipurth2012}. Since more massive stars boast a higher companion frequency, this could therefore lead to a high occurrence of wide, extreme systems. It is not yet possible to perform the statistically analysis of this doubly rare category but the quality and size of the sample size is rapidly expanding, in large part thanks to {\it Gaia} \citep[e.g.,][]{Oelkers2017}. Preliminary analyses already suggest that wide multiple systems likely are produced through more than one channel, further complicating the interpretation.


\section{Conclusions}
\label{sec:concl}

We have conducted a high contrast imaging survey of 169 intermediate-mass (1.75--4.5\,$M_\odot$) stars using the (non-coronagraphic) ShaneAO and (coronagraphic) GPI AO systems to search for companions corresponding to mass ratios in the $0.05 \lesssim q \lesssim 0.1$, a domain corresponding to the brown dwarf desert among solar-type stars, at separations ranging from $\approx$20 to $\approx$1000\,au. The sample consists of a mix of nearby field stars and candidate members of open clusters, after excluding known bright binaries that would negatively affect the survey sensitivity. We identified 24 candidate companions, 16 of which are newly reported here, including 8 detected after applying advanced PSF subtraction algorithms (RDI LOCI for the ShaneAO datasets, and cADI and KLIP for the GPI datasets). We obtained more than one epoch for all but 4 of the multiple systems detected in the ShaneAO datasets, confirming common proper motion for all companions. We also present $H$ band spectra for the 4 companions detected in GPI datasets without PSF subtraction, deriving spectral types that are broadly consistent with their brightness, supporting the hypothesis that they are physically associated to their host star. The 3 fainter GPI companions could be substellar companions but are much more likely to be background stars based on the predicted density of stars in the direction of these sources.

All of the likely and confirmed companions have moderate near-infrared contrast ratio ($\lesssim6.5\,$\,mag in the $H$ or $K$ band) even though our survey achieved a sensitivity down to 9--10\,mag after PSF subtraction. Assuming that the companions are coeval with their primary star, we inferred their mass from evolutionary models and derived masses that range from $\approx$0.3 to 2\,$M_\odot$, corresponding to mass ratios $q \geq 0.1$. The lack of companions detected in this survey in the $0.05 \lesssim q \lesssim 0.1$ range is statistically confirmed by modeling the distribution of companions with power laws as a function of companion mass (or mass ratio) and semi-major axis. In particular, we find 3$\sigma$ lower limits of $M_\mathrm{min} = 0.21\,M_\odot$ and $q_\mathrm{min} = 0.073$, providing new support for a lower cutoff to the distribution of stellar binaries. 

Taking into account the growing population of brown dwarf companions to intermediate-mass stars, our survey reveals a ``low-mass companion desert" visual binaries, which quantitatively matches the solar-type brown dwarf desert if it is considered as a mass ratio feature. This result therefore confirms that the mass ratio of binary systems is more informative that the absolute mass of either component. Furthermore, we conclude that this desert delineates two distinct formation regimes, with systems with $q \lesssim 0.02$ forming in disks and those with $q \gtrsim 0.07$ forming through cloud fragmentation across a broad stellar mass range. Correlations between the masses of prestellar cores, stars, and circumstellar disks combine to paint a picture where both cloud and disk fragmentations are largely scale-free processes, thus explaining that the mass ratio of multiple systems is the most relevant quantity to consider.

Using the observed distribution of mass ratios as a blueprint and scaling it as a function of stellar mass, we predict a significant occurrence rate of binaries made of a low-mass star and a brown dwarf, as well as a
steeply increasing number of giant planets towards lower masses around solar-type stars (at least down to $\approx 1\,M_\mathrm{Jup}$). Apart from a few exceptional cases, both types of systems are currently mostly out of reach of current instrumentation but future technological improvements should provide the sensitivity necessary to test both predictions.

\section*{Acknowledgements}

We are grateful to the staff at Lick and Gemini observatories for their support during the execution of our programs and to the GPIES team for support in planning, executing and reducing the GPI data presented here. GD, JTO, PK and BC acknowledge support from the NSF through grant 1413671, as well as from the UC Berkeley Undergraduate Research Apprentice Program. Based on observations obtained at the international Gemini Observatory, a program of NSF’s NOIRLab, which is managed by the Association of Universities for Research in Astronomy (AURA) under a cooperative agreement with the National Science Foundation on behalf of the Gemini Observatory partnership: the National Science Foundation (United States), National Research Council (Canada), Agencia Nacional de Investigaci\'{o}n y Desarrollo (Chile), Ministerio de Ciencia, Tecnolog\'{i}a e Innovaci\'{o}n (Argentina), Minist\'{e}rio da Ci\^{e}ncia, Tecnologia, Inova\c{c}\~{o}es e Comunica\c{c}\~{o}es (Brazil), and Korea Astronomy and Space Science Institute (Republic of Korea). This research has made use of data from the European Space Agency (ESA) mission Gaia, processed by the Gaia Data Processing and Analysis Consortium, of the Washington Double Star Catalog maintained at the U.S. Naval Observatory, of data obtained from or tools provided by the portal exoplanet.eu of The Extrasolar Planets Encyclopaedia, and of the SIMBAD database and Vizier catalogue access tool operated by the Centre de Donn\'ees Astronomiques de Strasbourg, France.

\section*{Data Availability}

The raw ShaneAO and GPI data used in this work are publicly available through the Lick and Gemini observatories' public archives, at {\tt https://mthamilton.ucolick.org/data/} and {\tt https://archive.gemini.edu/searchform}, respectively. All reduced data will be shared upon reasonable request to the corresponding author.



\bibliographystyle{mnras}




\appendix

\section{Rejected targets}
\label{sec:rejected}

A few of the targets that were observed during the course of this survey were later rejected from the statistical analysis, because their {\it Gaia}-measured distance differed by more than 20\,au from the distance estimated by our evolutionary model fit (see Section\,\ref{subsec:sample}). This is typically because they are post-Main Sequence stars, whose properties are highly sensitive to small photometric errors, implying that their stellar properties are highly uncertain. This prevents us from using them in the analysis, and we do not consider them in our statistical analysis. However, we nonetheless used their images in the PSF subtraction process as the quality of the AO correction is independent of the physical stellar properties. Among these seven objects, HIP\,61394 is the only one that possesses a close and bright companion that is detected in our (ShaneAO) observations. It lies at a separation of 0\farcs37, has a flux ratio of $\Delta K \approx$1.4\,mag, and was previously known in the WDS catalog.

\begin{table*}
	\centering
	\caption{Targets observed during this survey but rejected from the analysis due to inconsistencies between the distances listed in {\it Gaia} (third column) and derived from our fit to stellar properties (fifth column, see Section\,\ref{subsec:sample}).}
	\label{tab:rejected}
	\begin{tabular}{cccccccc} 
		\hline
 Target & Sp.T. & $D_{Gaia}$ & $M_\star$ & $D_{MCMC}$ & $t_{MCMC}$ & Instrument & Obs. Date \\
 & & (pc) & ($M_\odot$) & (pc) & (Myr) & & (UT) \\
		\hline
HD 58260  &  B3III  & $742 \pm 24$ & 5.40 $^{+0.65}_{-0.25 }$ &  689.0 $^{+24.9}_{-19.0 }$ &  89 $^{+13}_{-18 }$ & GPI & 2015-12-26  \\ 
HD\,75137  & A0Vn & $254 \pm 124$ & 3.05 $^{+0.17}_{-0.14}$ &  99.7 $^{+10.1}_{-8.7}$ &  351 $^{+35}_{-35}$ & ShaneAO & 2016-03-27 \\ 
HIP\,61394  & A0IV & $433 \pm 176$ & 2.80 $^{+0.11}_{-0.12}$ &  96.8 $^{+5.0}_{-4.5}$ &  424 $^{+40}_{-32}$ & ShaneAO & 2016-03-27 \\ 
HR\,6719  & B2IV & $688 \pm 27$ & 5.39 $^{+0.79}_{-0.07}$ &  502.0 $^{+94.8}_{-7.4}$ &  90 $^{+3}_{-18}$ & ShaneAO & 2015-06-07 \\ 
HR\,6873  & B2Vne & $417 \pm 14$ & 3.66 $^{+0.45}_{-0.05}$ &  327.9 $^{+5.8}_{-12.9}$ &  247 $^{+10}_{-68}$ & ShaneAO & 2015-06-06 \\ 
ups01 Pup  &  B3Ve  & $222 \pm 13$ & 4.06 $^{+0.14}_{-0.40 }$ &  159.5 $^{+4.6}_{-3.9 }$ &  185 $^{+61}_{-16 }$ & GPI & 2015-12-26 \\ 
V986\,Oph  & O9.7IIn & $1110 \pm 91$ & 5.00 $^{+0.06}_{-0.06}$ &  408.7 $^{+6.7}_{-5.8}$ &  108 $^{+3}_{-3}$ & ShaneAO & 2015-06-07 \\
        \hline
	\end{tabular}
\end{table*}

\section{ShaneAO Astrometric Calibration}
\label{sec:shaneao_calib}

To determine the astrometric calibration of ShARCS, which was unknown at the time of our observations, we included observations of several known wide binaries during each ShaneAO observing run. These systems were selected from the WDS, based on their brightness, separation and the expected precision of their predicted position over the course of this project. The orbit quality grades, as listed in the WDS catalog, range from 2 ("good") to 4 ("preliminary") and 5 ("indeterminate"). The latter two categories, while indicating poorly determined orbits, do not preclude precise prediction of the position of the binary over timescale of a few years, as these systems have orbits that typically are in the 1000--4000\,yr range and their relative position is well predicted over spans of just a few years. For each system, we obtained a short sequence of typically 10 unsaturated frames split over two detector positions. Overall, we obtained 22 observations of calibration binaries, representing 14 distinct systems, spread over 7 observing runs, as listed in Table\,\ref{tab:calbin}.

We measured the position of both components in each calibration binary using the centroid method. The resulting offsets in pixel were converted to an estimate of the plate scale and absolute orientation of the detector. Our observations, using only a few positions on the detector and a small fraction of the full field of view, are inappropriate to estimate the amplitude of camera distortion effects and we assume that these average out given the diversity of binary separations and position angles. Results of this process are indicated in Table\,\ref{tab:calbin} and illustrated in Fig.\,\ref{fig:astrom_calib}. We found no significant difference between the average plate scale and orientation of the detector from run to run, as indicated by the repeated observations of eight systems throughout the project. We thus proceeded to compute a straight average over all observed systems, excluding only WDS\,09357+3549 whose published orbit is clearly inconsistent with our measurement in October 2017. We use the standard deviation of the mean over all systems as our estimate of the astrometric calibration uncertainties, since the scatter between systems is much larger than the measurement uncertainties and likely stems from slight offsets relative to the published orbits. The results are consistent with those obtained in a separate program that aims at solving for both astrometric properties and camera distortion (Mark Ammons, priv. comm.).

\begin{figure}
\centering
\includegraphics[width=0.8\columnwidth]{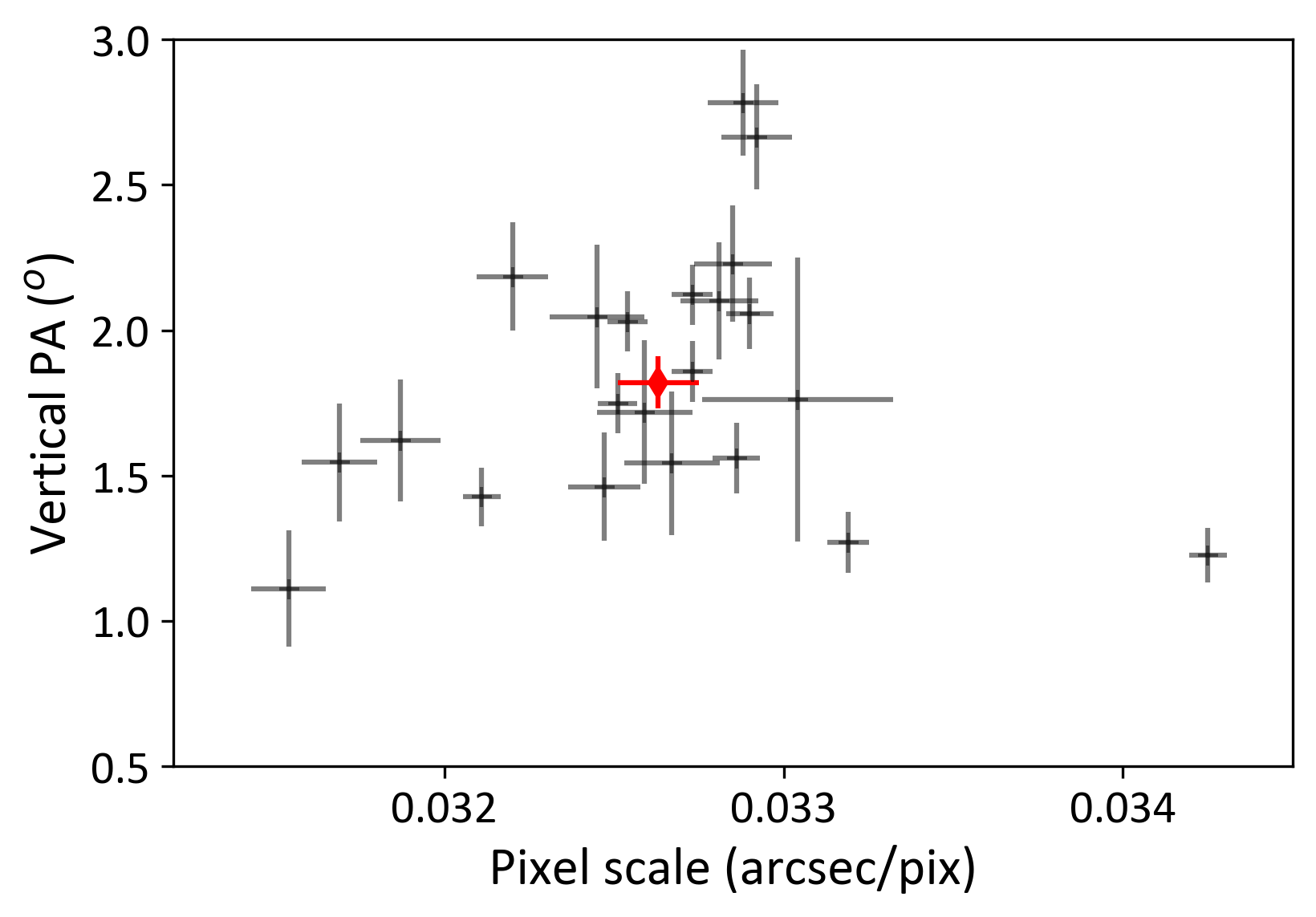}\\
\caption{ShaneAO pixel scale and absolute orientation of the vertical axis (measured East from North) as determined from our observations of calibration binaries. The red diamond represents the weighted average and standard deviation of the mean across the entire sample (except the outlier WDS\,09357+3549) on both quantities. \label{fig:astrom_calib}}
\end{figure}

\begin{table}
	\centering
	\caption{Astrometric binaries observed with ShARCS. The second column indicate the "orbit grade" listed in the WDS catalog, whereas the fifth column represents the position angle of the detector's Y axis, measured E from N, based on each system, as well as their weighted average (last row). The system labeled with a $\dagger$ symbol is a clear outlier and is excluded from the final average.}
	\label{tab:calbin}
	\begin{tabular}{ccccc} 
		\hline
 WDS & Grade & Obs. Date & Pixel Scale & PA$_{vert}$ \\
 & & (UT) & (mas/pix) & (\degr) \\
		\hline
00063+5826 & 2 & 2016-09-20 & 33.04 & 1.76 \\
05364+2200 & 5 & 2016-09-19 & 32.88 & 2.78 \\
 & & 2017-10-05 & 32.92 & 2.67 \\
06546+1311 & 4 & 2017-10-04 & 32.11 & 1.43 \\
09357+3549$^\dagger$ & 5 & 2017-10-04 & 29.52 & -8.16 \\
13550+0804 & 5 & 2016-03-26 & 31.69 & 1.55 \\
 & & 2018-05-27 & 31.54 & 1.11 \\
14336+3535 & 5 & 2015-06-06 & 32.59 & 1.72 \\
 & & 2018-05-27 & 32.67 & 1.54 \\
15348+1032 & 4 & 2015-06-05 & 32.20 & 2.19 \\
 & & 2015-07-04 & 32.47 & 1.46 \\
16147+3352 & 4 & 2016-03-27 & 32.54 & 2.03 \\
 & & 2017-05-16 & 32.51 & 1.75 \\
18239+5848 & 5 & 2015-07-04 & 32.81 & 2.10 \\
 & & 2018-05-27 & 32.85 & 2.23 \\
19121+4951 & 5 & 2016-09-19 & 32.73 & 1.86 \\
 & & 2017-05-11 & 33.19 & 1.27 \\
 & & 2017-10-04 & 32.73 & 2.12 \\
19464+3344 & 2 & 2015-06-05 & 32.45 & 2.05 \\
20014+1045 & 5 & 2015-06-06 & 31.87 & 1.62 \\
20462+1554 & 4 & 2015-06-06 & 32.90 & 2.06 \\
 & & 2015-07-04 & 32.86 & 1.56 \\
22038+6438 & 5 & 2017-10-05 & 34.25 & 1.23 \\
\hline
Average & & & 32.63$\pm$0.12 & 1.82$\pm$0.09 \\
        \hline
	\end{tabular}
\end{table}

\section{Model Fitting to Complementary Subsamples}
\label{sec:model_subsamples}

For the main part of this study, we have combined the GPI and ShaneAO samples, irrespective of the nature of the target. Here we explore how splitting the sample between field stars and open cluster members (see Section\,\ref{subsec:sample}) affect our results. Fig.\,\ref{fig:tongueplots_cluster_field} presents the corresponding completeness plots. A single ShaneAO companion (the most distant one) is associated with a cluster member. Because the number of ShaneAO-observed cluster member is small, performing the fit on these two subsamples hardly changes the model fitting presented in Section\,\ref{subsec:model_distrib}. Furthermore, as it is likely that at least a few candidate cluster members are actually unrelated to their parent cluster, we decided to not perform the full statistical analysis based on this split of the full sample.

\begin{figure*}
\includegraphics[width=0.8\textwidth]{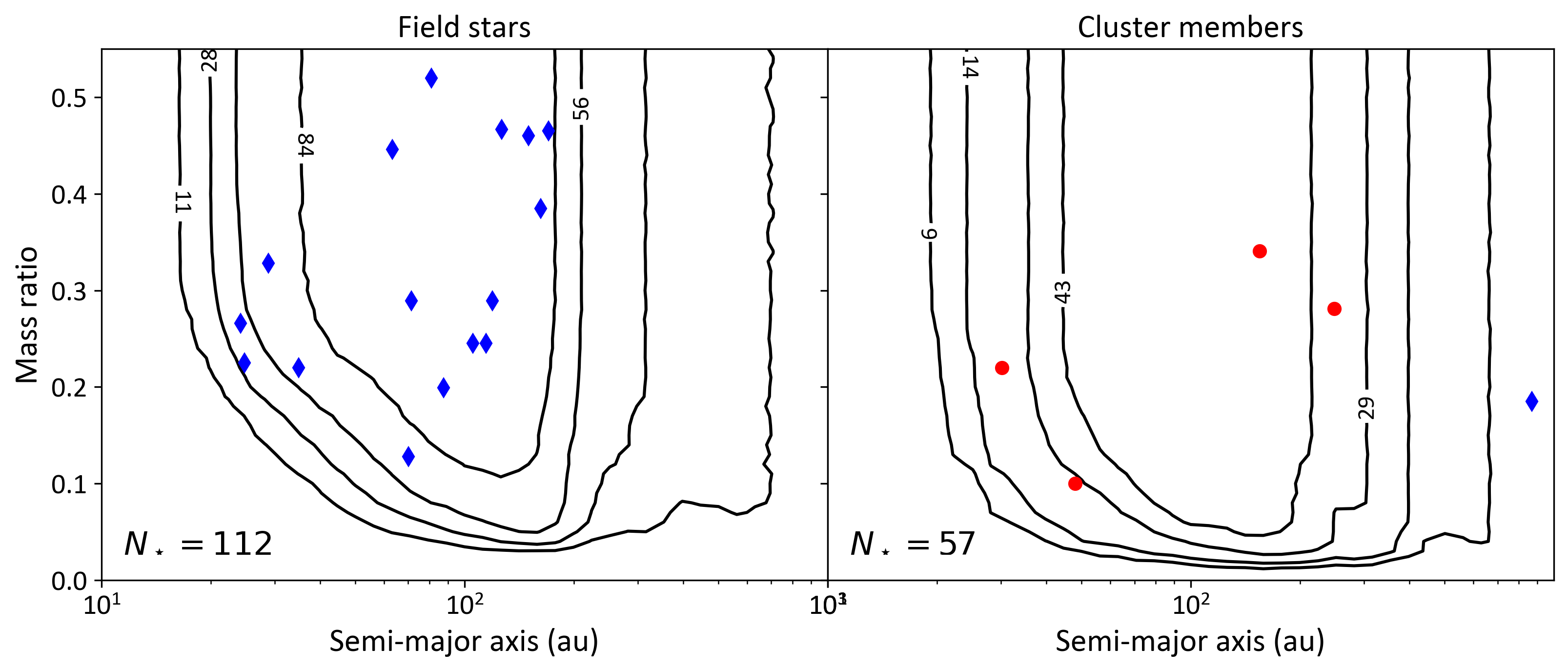}\\
\caption{Mass ratio survey completeness for the field and cluster subsamples (left and right column, respectively). Blue diamonds and red circles represent ShaneAO- and GPI-detected companions, respectively.  \label{fig:tongueplots_cluster_field}}
\end{figure*}

Another way to split the sample consists in separating the higher- and lower-mass targets in the survey. We adopt the median primary mass, 2.55\,$M_\odot$ to split the sample; the resulting completeness maps are shown in Fig.\,\ref{fig:tongueplots_primmass}. Two third of the companions detected in this survey are associated with the lower-mass targets, suggesting a higher companion frequency. This is however in part due to the better sensitivity to low-mass companions, as these have a more favorable flux ratio. When performing the same model fitting as in Section\,\ref{subsec:model_distrib}, the difference in companion frequency are consistent within the 2$\sigma$ confidence level. Furthermore, the confidence intervals for $\beta$, $\alpha_\mathrm{m}$, and $\alpha_\mathrm{q}$ for the low- and high-mass subsamples are statistically consistent with each other. Finally, the lower limits on $M_\mathrm{min}$ and $q_\mathrm{min}$ for the low-mass subsample are consistent with those of the full sample fit, whereas the high-mass subsample contains too few companions to provide meaningful constraints on these parameters.

\begin{figure*}
\includegraphics[width=0.8\textwidth]{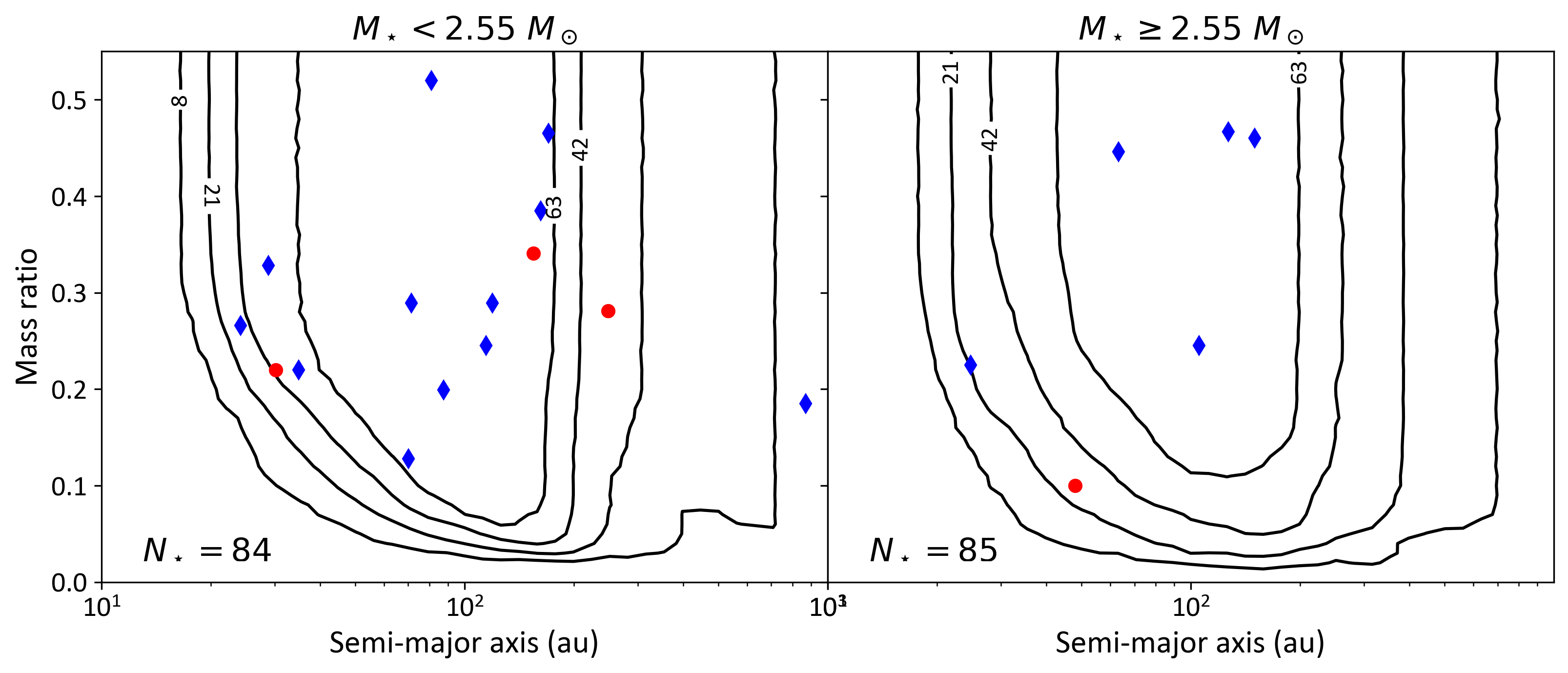}\\
\caption{Mass ratio survey completeness for stars above or below the median primary mass in the survey (left and right column, respectively). Blue diamonds and red circles represent ShaneAO- and GPI-detected companions, respectively. \label{fig:tongueplots_primmass}}
\end{figure*}

\section{Extreme Binary Systems From the Literature}
\label{sec:extreme_lit}

In addition to collating the companions detected in large multiplicity surveys, we have also performed a thorough search for ``extreme" binary systems. Here we describe the methods we used to assemble this list, which is presented in Table\,\ref{tab:extreme_lit}. 

The primary source to build this list is the Exoplanet Encyclopedia\footnote{http://exoplanet.eu/}, from which we selected companions detected by imaging methods, with mass above 10\,$M_\mathrm{Jup}$, and with separation in the 10--1500\,au range. In some cases, the mass of the primary star was missing and it was supplemented from literature on the object. We complemented this list with companions listed in \cite{Deacon2014}, \cite{Bowler2016} and \cite{Vigan2021} that match the same criteria. Finally, we added all the companions not included in these catalogs that we were aware of. 

We adopted the primary and companion masses from discovery papers, except in cases where a significantly different (and generally more precise) estimate was published in a subsequent analysis. The mass of substellar companions is often presented with a large uncertainty that stems from uncertainties in the system's age, and/or as an acknowledgement of significant model uncertainties. When this applied, we computed the geometric mean of the lower and maximum mass listed in the relevant study and used this as a representative mass of the companion. Assuming that these errors are random in nature, with roughly as many under- and over-estimate, the results of our analysis should not be significantly affected by this issue. Finally, in the case of hierarchical triple systems, we adopt the higher mass of the closer pair as the primary or companion mass, effectively ignoring one of the components of the system.

\begin{table*}
	\centering
	\caption{Previously known extreme binary systems. References for the companion mass (fifth and tenth columns): 1) \citet{Mamajek2010}; 2) \citet{Lagrange2010}; 3) \citet{der14}; 4) \citet{Hinkley2010}; 5) \citet{Neuhauser2011}; 6) \citet{Hinkley2013}; 7) \citet{Nielsen2012}; 8) \citet{Claudi2019}; 9) \citet{Musso2019}; 10) \citet{Cheetham2018b}; 11) \citet{hin15}; 12) \citet{Wagner2020}; 13) \citet{Lafreniere2011}; 14) \citet{Janson2019}; 15) \citet{maw15}; 16) \citet{Adam2013}; 17) \citet{Huelamo2010}; 18) Exoplanets Encyclopedia; 19) \citet{Deacon2014}; 20) \citet{Johnson2017}; 21) \citet{Cheetham2018a}; 22) \citet{Bonavita2022}; 23) \citet{Potter2002}; 24) \citet{Bowler2016}. Systems where the low mass companion is exterior to a close stellar binary are indicated with a $\dagger$ symbol.
	}
	\label{tab:extreme_lit}
	\begin{tabular}{cccccccccccc} 
		\hline
System & $M_\mathrm{prim}$ & $M_\mathrm{comp}$ & $q$ & Sep. & Ref. & System & $M_\mathrm{prim}$ & $M_\mathrm{comp}$ & $q$ & Sep. & Ref. \\
 & $(M_\odot)$ & $(M_\odot)$ & & (au) & & & $(M_\odot)$ & $(M_\odot)$ & & (au) & \\
		\hline
\multicolumn{12}{c}{Intermediate-Mass Stars}\\
\hline
$\alpha$\,UMa & 1.8 & 0.3 & 0.167 & 28 & 1 & HIP\,71724 & 3.41 & 0.178 & 0.052 & 16 & 11 \\
$\beta$\,Pic & 1.73 & 0.012 & 0.007 & 10 & 2 & HIP\,73990\,AB & 1.72 & 0.032 & 0.019 & 20 & 11 \\
$\zeta$\,Del & 2.5 & 0.054 & 0.022 & 910 & 3 & HIP\,73990\,AC & 1.72 & 0.034 & 0.020 & 32 & 11 \\
$\zeta$\,Vir & 2.04 & 0.165 & 0.081 & 40 & 4 & HIP\,75056 & 1.92 & 0.024 & 0.013 & 16 & 12 \\
$\eta$\,Tel & 2.2 & 0.032 & 0.015 & 200 & 5 & HIP\,78233 & 1.67 & 0.10 & 0.060 & 19 & 11 \\
$\kappa$\,And & 2.8 & 0.049 & 0.018 & 55 & 6 & HIP\,78530 & 2.5 & 0.022 & 0.009 & 710 & 13 \\ 
HD\,1160\,AB & 2.2 & 0.033 & 0.015 & 81 & 7 & HIP\,79098$^\dagger$ & 2.5 & 0.020 & 0.008 & 345 & 14 \\ 
HD\,1160\,AC & 2.2 & 0.21 & 0.095 & 533 & 7 & HIP\,79124 & 2.48 & 0.17 & 0.069 & 22 & 11 \\
HD\,142527 & 2.2 & 0.12 & 0.055 & 13 & 8 & HR\,3549 & 2.32 & 0.037 & 0.016 & 80 & 15 \\
HD\,193571 & 2.2 & 0.35 & 0.159 & 11 & 9 & HR\,3672 & 3.5 & 0.39 & 0.111 & 436 & 16 \\
HIP\,64892 & 2.35 & 0.033 & 0.014 & 159 & 10 & HR\,6037 & 1.8 & 0.059 & 0.033 & 366 & 17 \\
\hline
\multicolumn{12}{c}{Solar-Type Stars}\\
\hline
1RXS\,1609 & 0.73 & 0.014 & 0.019 & 331 & 18 & HD\,130948 & 1.0 & 0.054 & 0.054 & 47 & 23 \\
2M\,2236+4751 & 0.6 & 0.013 & 0.022 & 233 & 18 & HD\,203030 & 0.97 & 0.011 & 0.011 & 488 & 18 \\
AB\,Pic & 0.97 & 0.014 & 0.014 & 275 & 18 & HD\,206893 & 1.24 & 0.018 & 0.015 & 10 & 18 \\
CT\,Cha & 0.7 & 0.017 & 0.024 & 441 & 18 & HD\,284149$^\dagger$ & 1.14 & 0.032 & 0.028 & 398 & 18 \\
GJ\,758 & 0.97 & 0.038 & 0.039 & 55 & 18 & HII\,1348$^\dagger$ & 0.65 & 0.056 & 0.086 & 132 & 19 \\
GJ\,1048 & 0.77 & 0.064 & 0.083 & 251 & 19 & HIP\,21152 & 1.44 & 0.026 & 0.018 & 18 & 22 \\
Gl\,337$^\dagger$ & 0.90 & 0.055 & 0.061 & 881 & 19 & HIP\,63734 & 1.21 & 0.011 & 0.009 & 30 & 18 \\
Gl\,618.1 & 0.67 & 0.069 & 0.103 & 1090 & 19 & HIP\,74865 & 1.42 & 0.043 & 0.030 & 23 & 11 \\
GQ\,Lup & 0.7 & 0.013 & 0.019 & 103 & 18 &  HIP\,112581 & 1.09 & 0.094 & 0.086 & 28 & 22 \\
GSC\,08047-00232 & 0.8 & 0.025 & 0.031 & 282 & 18 & HN\,Peg & 1.0 & 0.016 & 0.016 & 795 & 18 \\
GSC\,06214+210 & 0.6 & 0.017 & 0.028 & 320 & 18 & HR\,2562 & 1.3 & 0.026 & 0.020 & 20 & 18 \\
HD\,984 & 1.2 & 0.057 & 0.048 & 10 & 20 & PZ\,Tel & 1.13 & 0.064 & 0.057 & 25 & 18 \\
HD\,3651 & 0.79 & 0.053 & 0.067 & 473 & 18 & Ross\,458$^\dagger$ & 0.6 & 0.011 & 0.018 & 120 & 18 \\
HD\,4113 & 1.05 & 0.036 & 0.034 & 22 & 21 & ROXs\,12 & 0.87 & 0.016 & 0.018 & 210 & 18 \\
HD\,19467 & 0.95 & 0.052 & 0.055 & 51 & 18 & ROXs\,42\,B$^\dagger$ & 0.89 & 0.010 & 0.011 & 150 & 24 \\
HD\,33632 & 1.03 & 0.050 & 0.049 & 21 & 18 & SR\,12$^\dagger$ & 1.0 & 0.013 & 0.013 & 1088 & 24 \\
HD\,60584 & 1.35 & 0.015 & 0.011 & 17 & 22 & TYC\,8998-760-1 & 1.0 & 0.014 & 0.014 & 162 & 18 \\
HD\,65216 & 0.94 & 0.094 & 0.010 & 253 & 19 & USco\,1602-2401 & 1.34 & 0.044 & 0.033 & 1020 & 19 \\
HD\,106906$^\dagger$ & 1.5 & 0.011 & 0.007 & 654 & 18 & USco\,1610-1913 & 0.88 & 0.021 & 0.024 & 841 & 19 \\
\hline
	\end{tabular}
\end{table*}

\bsp	
\label{lastpage}
\end{document}